\documentclass[prx, preprint, superscriptaddress]{revtex4-1}  

\usepackage{graphicx}  
\usepackage{bm} 
\usepackage{amssymb}  
\usepackage{amsmath}
\usepackage{mhchem}
\usepackage{mathptmx}
\usepackage{siunitx}
\usepackage{afterpage}
\usepackage{color}
\begin{document}
\widetext


\title{Tunable layered-magnetism-assisted magneto-Raman effect in a two-dimensional magnet \ce{CrI3}}


\author{Wencan Jin}
\altaffiliation{current affiliation: Department of Physics, Auburn University, 380 Duncan Drive, Auburn, AL 36849, USA}
\affiliation{Department of Physics, University of Michigan, 450 Church Street, Ann Arbor, Michigan 48109, USA}

\author{Zhipeng Ye}
\affiliation{Department of Electrical and Computer Engineering, 910 Boston Avenue, Texas Tech University, Lubbock, Texas 79409, USA}

\author{Xiangpeng Luo}
\affiliation{Department of Physics, University of Michigan, 450 Church Street, Ann Arbor, Michigan 48109, USA}

\author{Bowen Yang}
\affiliation{Institute for Quantum Computing, Department of Chemistry, and Department of Physics and Astronomy, University of Waterloo, Waterloo, 200 University Ave W, Ontario N2L 3G1, Canada}

\author{Gaihua Ye}
\affiliation{Department of Electrical and Computer Engineering, 910 Boston Avenue, Texas Tech University, Lubbock, Texas 79409, USA}

\author{Fangzhou Yin}
\affiliation{Institute for Quantum Computing, Department of Chemistry, and Department of Physics and Astronomy, University of Waterloo, Waterloo, 200 University Ave W, Ontario N2L 3G1, Canada}

\author{Hyun Ho Kim}
\altaffiliation{current affiliation: School of Materials Science and Engineering, Kumoh National Institute of Technology, Gumi, Gyeongbuk 39177, Korea}
\affiliation{Institute for Quantum Computing, Department of Chemistry, and Department of Physics and Astronomy, University of Waterloo, Waterloo, 200 University Ave W, Ontario N2L 3G1, Canada}

\author{Laura Rojas}
\affiliation{Department of Electrical and Computer Engineering, 910 Boston Avenue, Texas Tech University, Lubbock, Texas 79409, USA}

\author{Shangjie Tian}
\affiliation{Department of Physics and Beijing Key Laboratory of Opto-electronic Functional Materials \& Micro-nano Devices, Renmin University of China, Beijing 100872 China}

\author{Yang Fu}
\affiliation{Department of Physics and Beijing Key Laboratory of Opto-electronic Functional Materials \& Micro-nano Devices, Renmin University of China, Beijing 100872 China}

\author{Shaohua Yan}
\affiliation{Department of Physics and Beijing Key Laboratory of Opto-electronic Functional Materials \& Micro-nano Devices, Renmin University of China, Beijing 100872 China}

\author{Hechang Lei}
\affiliation{Department of Physics and Beijing Key Laboratory of Opto-electronic Functional Materials \& Micro-nano Devices, Renmin University of China, Beijing 100872 China}

\author{Kai Sun}
\affiliation{Department of Physics, University of Michigan, 450 Church Street, Ann Arbor, Michigan 48109, USA}

\author{Adam W. Tsen}
\affiliation{Institute for Quantum Computing, Department of Chemistry, and Department of Physics and Astronomy, University of Waterloo, Waterloo, 200 University Ave W, Ontario N2L 3G1, Canada}

\author{Rui He}
\email{rui.he@ttu.edu}
\affiliation{Department of Electrical and Computer Engineering, 910 Boston Avenue, Texas Tech University, Lubbock, Texas 79409, USA}

\author{Liuyan Zhao} 
\email{lyzhao@umich.edu}
\affiliation{Department of Physics, University of Michigan, 450 Church Street, Ann Arbor, Michigan 48109, USA}



\begin{abstract}
\noindent We use a combination of polarized Raman spectroscopy experiment and model magnetism-phonon coupling calculations to study the rich magneto-Raman effect in the two-dimensional (2D) magnet \ce{CrI3}. We reveal a novel layered-magnetism-assisted phonon scattering mechanism below the magnetic onset temperature, whose Raman excitation breaks time-reversal symmetry, has an antisymmetric Raman tensor, and follows the magnetic phase transitions across critical magnetic fields, on top of the presence of the conventional phonon scattering with symmetric Raman tensors in $N$-layer \ce{CrI3}. We resolve in data and by calculations that the $1^\mathrm{st}$-order $A_\mathrm{g}$ phonon of monolayer splits into a $N$-fold multiplet in $N$-layer \ce{CrI3} due to the interlayer coupling ($N \ge 2$) and that the phonons with the multiple show distinct magnetic field dependence because of their different layered-magnetism-phonon coupling. We further find that such a layered-magnetism-phonon coupled Raman scattering mechanism extends beyond $1^\mathrm{st}$-order to higher-order multi-phonon scattering processes. Our results on magneto-Raman effect of the $1^\mathrm{st}$-order phonons in the multiplet and the higher-order multi-phonons in $N$-layer \ce{CrI3} demonstrate the rich and strong behavior of emergent magneto-optical effects in 2D magnets and underlines the unique opportunities of new spin-phonon physics in van der Waals layered magnets.
\end{abstract}
\maketitle


Two-dimensional (2D) \ce{CrI3} of few-layer form features a layered-antiferromagnetic (AFM) order where the spins align along the out-of-plane direction ferromagnetically within each layer and antiferromagnetically between adjacent layers \cite{RN1,RN2,RN3,RN4, RN5}. It undergoes a layered-AFM to FM phase transition upon applying a moderate magnetic field \cite{RN1, RN2, RN3, RN4, RN5, RN6, RN7}, or electric field \cite{RN8,RN9, RN10A}, or electrostatic doping \cite{RN10}, or hydrostatic pressure \cite{RN11, RN12}. The strong coupling between spin and charge degrees of freedom in 2D \ce{CrI3} allows magneto-optical effects manifested in a variety of ways including large magneto-optical Kerr effect \cite{RN5} and magnetic circular dichroism \cite{RN8, RN9, RN10A, RN10, RN11, RN12}, spontaneous helical photoluminescence \cite{RN13}, giant nonreciprocal second harmonic generation \cite{RN14}, and anomalous magneto-optical Raman effect \cite{RN7, RN15, RN16, RN17, RN18}. All of these magneto-optical effects can be tuned across the layered-AFM to FM phase transition, making 2D \ce{CrI3} a promising candidate for applications in magnetic sensors, optical modulation, and data storage.

Among all magneto-optical effects in \ce{CrI3}, magneto-optical Raman effect is of particular interest for two reasons. First, among all known magnets, the largest magnetism-induced optical rotation is observed for the linearly polarized, inelastically scattered light off the $A_\mathrm{g}$ phonon mode ($\sim$129 $\mathrm{cm}^{-1}$) in the FM phase of \ce{CrI3} \cite{RN15}. Second, different phonon modes exhibit distinct magneto-optical behavior that the $A_\mathrm{g}$ mode emerges whereas its neighboring antisymmetric mode ($\sim$127 $\mathrm{cm}^{-1}$) disappears in the linear crossed polarization channel across the layered-AFM to FM transition  \cite{RN7, RN15, RN16, RN17}. However, the physical origin of magneto-optical Raman effect remains elusive with diverse proposals ranging from Davydov-split for bilayer \ce{CrI3} \cite{RN15, RN17}, zone-folded phonon for few-layer \ce{CrI3} \cite{RN16}, and coupled magnetism-phonon scattering for bulk \ce{CrI3} \cite{RN7}, none of which can be trivially generalized to explain \ce{CrI3} of arbitrary thickness.

In this work, we carefully examine the magneto-optical Raman effect for $N$-layer \ce{CrI3} ($N = 1-4$) by performing polarization, temperature, and magnetic field dependent micro-Raman spectroscopy measurements and unambiguously identify the layered-magnetism-assisted phonon scattering as the origin applicable for \ce{CrI3} of any thickness. $N$-layer \ce{CrI3} flakes were exfoliated from high-quality \ce{CrI3} single crystals, sandwiched between hexagonal boron nitride (hBN) thin flakes, and then placed onto \ce{SiO2}/Si substrates inside a high-purity ($> 99.999\%$) nitrogen-filled glovebox. Micro-Raman spectroscopy measurements in the backscattering geometry were carried out with a 633 nm excitation laser resonant with the charge-transfer transition \cite{RN13}, inside a vacuum cryostat at a base pressure lower than $7\times10^{-7}$ mbar, and under an out-of-plane magnetic field ($B_\bot$) up to 2.2 T.

We start with resolving in $N$-layer \ce{CrI3} the interlayer coupling-induced split of the $A_\mathrm{g}$ mode of monolayer \ce{CrI3} \cite{RN19, RN20}. Figure 1A shows Raman spectra in both parallel and crossed linear polarization channels taken at $T$ = 10 K and $B_\bot$ = 0 T on 1 -- 4L \ce{CrI3} (see full-range spectra in both channels and comparison to off-resonance 532 nm excitations in \textit{SM} Section I, Figures S1 and S2, respectively). It has been established for \ce{CrI3} that the modes in the crossed channel in Fig. 1A correspond to antisymmetric Raman tensor ($R_\mathrm{AS}$) whereas those in the parallel channel are for symmetric Raman tensor of $A_\mathrm{g}$ symmetry ($R_\mathrm{S}$) \cite{RN18}. We highlight three key observations that have not been reported in previous work \cite{RN7, RN15, RN16, RN17, RN18, RN21, RN22} and summarize them in Fig. 1B with fitted mode frequencies vs. $N$. First, the number of modes increases proportional to the number of layers (with an exception for $N = 3$ that is explained in \textit{SM} Section II). Second, the highest frequency remains constant while the lowest frequency decreases with increasing $N$, leading to a greater frequency separation between them. Third, the parallel and crossed channels show modes of the same frequencies for odd $N$ whereas they select modes with distinct frequencies for even $N$. To interpret the $A_\mathrm{g}$ mode splitting, we take a simple linear chain model of $N$-layer \ce{CrI3}, as introduced in few-layer transition metal dichalcogenides \cite{RN23, RN24, RN25, RN26}, \
\begin{center}
$H=H_0+\frac{1}{2}\sum_{i=2}^{N}k(u_{i-1}-u_i)^2$, with $H_0=\sum_{i=1}^{N}(\frac{1}{2}m\dot{u}_i^2+\frac{1}{2}k_0u_i^2)$ \
\end{center}

\noindent where $H_0$ represents the original $A_\mathrm{g}$ mode at frequency $\omega_0=\sqrt{k_0/m}$ within individual layers, $u_i$ represents the displacement field in the $i^\mathrm{th}$ layer, and $k$ stands for the coupling constant between adjacent layers, equivalent to a coupling frequency $\omega=\sqrt{k/m}$. Diagonalizing $H$ leads to $N$ nondegenerate eigenfrequencies $\Omega_i (\omega_0,\omega)$ and their corresponding eigenmodes $\vec{U}_i (i = 1, 2, \cdots, N)$ (\textit{i.e.}, Davydov-splitting), with $i = 1$ being the highest frequency mode and $i = N$ being the lowest frequency mode. By choosing $\omega_0=129.10\pm0.10$ $\mathrm{cm}^{-1}$ and $\omega=15.98\pm0.55 \mathrm{cm}^{-1}$, the calculated frequencies $\Omega_i$ (solid lines with open diamonds) match well with all the experimental values (ruby filled squares and royal filled circles), as highlighted by the fan-diagram in Fig 1B. See detailed calculations of $\Omega_i$ and $\vec{U}_i$ in \textit{SM} Section II, Table S1 and S2.

Because $N$-layer \ce{CrI3} is structurally centrosymmetric, its $N$ calculated eigenmodes have alternating parities, with the highest-frequency mode always parity-even as a result of equal, in-phase atomic displacement between layers (\textit{i.e.}, $\vec{U}_1=(1,1,\cdots,1)/\sqrt{N})$. In the parallel channel where only modes with even parity and symmetric Raman tensor $R_\mathrm{S}$ can be detected, we expect to see every other mode starting with the highest frequency one ($i = 1, 3, 5, \cdots$). This expectation is indeed consistent with our data that $U_1$ for $N = 1$ and 2 and $U_{1,3}$ for $N = 3$ and 4 are observed in the linear parallel channel in Fig. 1A and 1B. In contrary to the structure of $N$-layer \ce{CrI3}, the layered-AFM order is centrosymmetric for odd $N$ and non-centrosymmetric for even $N$. Therefore, it should couple to parity-even phonon modes for odd $N$ and parity-odd phonon modes for even $N$ to make the coupled layered-AFM-phonon entity parity-even and thus Raman-active. Due to the broken time-reversal symmetry from the magnetism, this layer-AFM-assisted phonon scattering corresponds to anti-symmetric Raman tensor $R_\mathrm{AS}$ and can only appear in the linear crossed channel. We anticipate observing in the crossed channel every other mode from the lowest-frequency one ($i = N, N-2, N-4, \cdots$), because $\vec{U}_N$ always has the same parity as the layered-AFM for any $N$. The result in Fig. 1A and 1B corroborates with this prediction that $U_1$ for $N = 1$, $U_2$ for $N = 2$, $U_{3,1}$ for $N = 3$, and $U_{4,2}$ for $N = 4$ are present in the linear crossed channel. The coupling efficiency between a phonon mode and the layered-AFM order can be evaluated by the projection $\vec{U}_i\cdot \vec{M}$ of the eigenmode vector $\vec{U}_i$ onto the pseudo-vector (\textit{i.e.}, axial vector) for the layered-AFM $\vec{M}=(1,-1,\cdots,(-1)^{N-1})$ with +1 for spin up and -1 for spin down in a single layer. In particular, for any $N > 1$, the lowest-frequency mode $\vec{U}_N$ features out-of-phase atomic displacement between adjacent layers and matches best the pattern of alternating spin orientations in the layered-AFM state (Fig. 1C), yielding the strongest coupling strength ($\vec{U}_i\cdot \vec{M}$) and thus the most intense signal among the modes in the linear crossed channel (Fig. 1A). See the computed $\vec{U}_i\cdot \vec{M}$ in the layered-AFM state of 1-4L \ce{CrI3} in \textit{SM}, Table S3. 

So far we have established the physical origin of the $N$-fold multiplet for $N$-layer \ce{CrI3} as a combined effect of Davydov-splitting and layered-AFM-phonon coupling, leading to the conventional phonons of $R_\mathrm{S}$ in the linear parallel channel \cite{RN23, RN24, RN25, RN26} and the layered-AFM-coupled phonons of $R_\mathrm{AS}$ in the linear crossed channel \cite{RN7, RN18}. We note that a magnetism-induced symmetric $E_\mathrm{g}$ phonon mode splitting was previously reported in 2D \ce{Cr2Ge2Te6} \cite{Tian2016}. Here, the structural and magnetic nature of modes in the linear parallel and crossed channel, respectively, is further supported by their distinct temperature dependence of modes in these two channels. Taking 2L \ce{CrI3} as an example, the linear crossed channel signal emerges below the magnetic transition temperature $T_\mathrm{C}$ = 45 K whereas the parallel channel signal is present above $T_\mathrm{C}$ and only increases slowly below $T_\mathrm{C}$, as illustrated by representative spectra taken at 70 K, 40 K, and 10 K in Fig. 2A. Such a behavior extends beyond the $1^\mathrm{st}$-order phonons (Fig. 2A, left) to the $2^\mathrm{nd}$, and $3^\mathrm{rd}$-order ones (Fig. 2A, middle and right, respectively). For all three orders, the temperature dependence of integrated intensity (I. I.) in the linear crossed channel fits well with an order parameter-like function $I_0+I\sqrt{T_\mathrm{C}-T}$ (royal curves in Fig. 2B), in contrast to those in the linear parallel channel following a smooth anharmonic decay behavior \cite{RN27} (ruby curves in Fig. 2B). As pictorially summarized in Fig. 2C, we propose that a multi-phonon process \cite{RN28, RN29} leads to conventional $1^\mathrm{st}$, $2^\mathrm{nd}$, and $3^\mathrm{rd}$-order phonon modes (ruby) in the linear parallel channel, and its coupling with the layered-AFM order results in the magnetic counterparts (royal) in the linear crossed channel.

We then proceed to explore the magnetic field dependence of the layered-magnetism-assisted phonon modes in $N$-layer \ce{CrI3}. From now on, we chose circularly polarized light to perform Raman measurements for preventing any artifacts from Faraday rotation of light passing through optical components in stray magnetic field. In this work, we focus on two representative thicknesses, 2L and 4L \ce{CrI3}, having one and two critical magnetic transitions, respectively. We note that the mechanism described below is applicable for arbitrary $N$-layer \ce{CrI3}.

The 2L \ce{CrI3} undergoes a layered-AFM to FM transition at a critical magnetic field $B_\mathrm{c}=\pm 0.6$ T \cite{RN5}. Figure 3A presents Raman spectra of $1^\mathrm{st}$-order modes taken at $B_\bot$ = 0 T and $\pm$ 1.4 T, below and above $B_\mathrm{c}$, respectively, at 10 K in both LL and RR channels, where LL(RR) stands for the polarization channel in which the incident and scattered light is left(right)-hand circularly polarized. At 0 T, both modes ($U_1$ and $U_2$) of 2L \ce{CrI3} are present in Raman spectra that are identical in LL and RR channels. At $\pm$1.4 T, only the high-frequency mode ($U_1$) survives, and it shows giant circular dichroism of $\pm 78\%$ ($\mathrm{\frac{I.I._{LL}-I.I._{RR}}{I.I._{LL}+I.I._{RR}}}$). See the comparison of selection rules between linear and circular polarization bases for 2L \ce{CrI3} in \textit{SM}, Table S5. The magnetic field dependence of $U_2$ integrated intensity clearly shows its disappearance at $B_\mathrm{c}$, whereas that of $U_1$ increases (decreases) abruptly in the LL (RR) channel at $B_\mathrm{c}$, as shown in Fig. 3B.  Figure 3E and 3F show the magnetic field dependence of the $2^\mathrm{nd}$ and $3^\mathrm{rd}$-order modes of 2L \ce{CrI3}. Both of them mimic the magnetic field dependence of $U_1$ with a reduction of circular dichroism above $B_\mathrm{c}$, $\pm 71\%$ for the $2^\mathrm{nd}$-order and $\pm 50\%$ for the $3^\mathrm{rd}$-order. This observation suggests the participation of $U_1$ in the $2^\mathrm{nd}$, and $3^\mathrm{rd}$-order multi-phonon process.

We refer to the layered-magnetism-phonon coupling that we have developed above to understand the magnetic field dependence of the two $1^\mathrm{st}$-order modes ($U_1$ and $U_2$) in 2L \ce{CrI3}. For each mode $\vec{U}_i$, its Raman tensor $R^i$ is composed of the conventional structural ($R_\mathrm{S}^i$) and the novel layered-magnetism-assisted magnetic ($R_\mathrm{AS}^i$) contributions, \textit{i.e.}, $R^i=R_\mathrm{S}^i+\lambda_\mathrm{i} R_\mathrm{AS}^i$, where $R_\mathrm{S}^i$ is magnetic field independent and is only present for parity-even modes, $R_\mathrm{AS}^i \propto \vec{U}_i\cdot \vec{M}$ reflects the magnetic origin and selects the zero-momentum component, and $\lambda_\mathrm{i}$ is ratio of the magnetic to structural contribution for the $i^\mathrm{th}$ mode that depends on microscopic parameters such as spin-orbit-coupling. Here, $\vec{M}$ changes from (1,-1) to $(\pm1,\pm1)$ across the layered-AFM to FM transition at $B_\mathrm{c}=\pm 0.6$ T with the fully polarized FM spin moments pointing upwards/downwards. Specifically, for the parity-even high-frequency mode of $\vec{U}_1=\frac{1}{\sqrt{2}}(1,1)$, $R_\mathrm{S}^1=\begin{pmatrix} a_1 & \\ & a_1\end{pmatrix}$ at all magnetic fields and $R_\mathrm{AS}^1=\vec{U}_i\cdot \vec{M} \begin{pmatrix} &-a_1 i \\ +a_1 i & \end{pmatrix}=0$ below $B_\mathrm{c}$ and $\begin{pmatrix} & \mp\sqrt{2} a_1 i \\ \pm\sqrt{2} a_1 i & \end{pmatrix}$ above $B_\mathrm{c}$ (Fig. 3C, top and bottom), whereas for the parity-odd low-frequency mode of $\vec{U}_2=\frac{1}{\sqrt{2}}(1,-1)$, $R_\mathrm{S}^2=0$ always and $R_\mathrm{AS}^2=\begin{pmatrix} &-\sqrt{2} a_2 i \\ +\sqrt{2} a_2 i & \end{pmatrix}$ below $B_\mathrm{c}$ and 0 otherwise (Fig. 3C, middle), where $i=\sqrt{-1}$ to account for the time-reversal symmetry and $a_\mathrm{i}$ stands for the Raman scattering strength of the $i^\mathrm{th}$ mode (see the calculated magnetic field dependence of $U_1$ and $U_2$ in \textit{SM}, Table S4 and S5). This model faithfully reproduces the magnetic field dependence of both $1^\mathrm{st}$-order modes of 2L \ce{CrI3} in LL and RR channels by tuning only $\lambda_\mathrm{i}$ and $a_\mathrm{i}$ (Fig. 3D). 

Different from 2L \ce{CrI3}, $N$-layer \ce{CrI3} ($N>2$) undergoes two spin-flip transitions with increasing $B_\bot$, one at $B_\mathrm{c1} =\pm0.7$ T for spins in surface layers and the other at $B_\mathrm{c2} = \pm1.6$ T for spins in interior layers \cite{RN1,RN2}. For simplicity but without losing any generality, we pick 4L \ce{CrI3} and focus on measurements with the upwards magnetic field and in the RR polarization channel. Figure 4A shows Raman spectra of $1^\mathrm{st}$-order modes taken at $B_\bot$ = 0 T, 1 T, and 2 T, below $B_\mathrm{c1}$, between $B_\mathrm{c1}$ and $B_\mathrm{c2}$, and above $B_\mathrm{c2}$, respectively, at 10 K and in the RR channel. At 0 T, we can only reliably resolve three out of 4-fold multiplet of 4L \ce{CrI3}, namely, $U_1$, $U_3$, and $U_4$ as fitted by the ruby, honey, and royal Lorentzian profiles, respectively. This is because $U_2$ is spectrally so close to $U_1$ but has a much weaker intensity (Fig. 1A), thus getting overwhelmed by the strong $U_1$ in the RR channel. We observe both $U_1$ and $U_4$ decrease subsequently at 1 T and 2 T to finite and zero intensity, respectively, whereas $U_3$ increases at 1 T and then decreases at 2 T. The detailed magnetic field dependence of the $U_1$, $U_3$, and $U_4$ integrated intensity is shown in Fig. 4B, displaying the contrasting trends of $U_3$ to $U_1$ and $U_4$, and those of the $2^\mathrm{nd}$ and $3^\mathrm{rd}$-order modes are shown in Fig. 4E and 4F, closely mimicking those of $U_1$.

We carry out a similar analysis as we have done for 2L \ce{CrI3} above, but add an additional intermediate magnetic phase $\vec{M}=(1,-1,1,1)$ between the layered-AFM of $(1,-1,1,-1)$ and the fully spin polarized FM of $(1,1,1,1)$. While the structural contribution ($R_\mathrm{S}^i$) is only present for parity-even modes, $U_1$ and $U_3$, and remains magnetic field independent, the layered-magnetism-coupled magnetic contribution ($R_\mathrm{AS}^i$) varies proportionally to $\vec{U}_i\cdot \vec{M}$ as $\vec{M}$ changes as a function of $B_\bot$. Figure 4C lists the modes that have finite coupling to every layered magnetic order and thus nonzero $R_\mathrm{AS}^i$, according to which the magnetic contribution of $U_1$ appears above $B_\mathrm{c1}$, that of $U_3$ emerges between $B_\mathrm{c1}$ and $B_\mathrm{c2}$, and $U_{2,4}$ present below $B_\mathrm{c2}$ (see the calculated magnetic field dependence of $U_1$, $U_2$, $U_3$, and $U_4$ in \textit{SM}, Table S6 and S7). By adjusting $\lambda_\mathrm{i}$ and $a_\mathrm{i}$, the ratio of the magnetic to structural contribution and the overall strength of the $i^\mathrm{th}$ mode, we successfully show the consistency between the experimental and calculated magnetic field dependence of $U_{1,3,4}$ and predict that of $U_2$ despite its invisibility in our experiment (Fig. 4D).

In conclusion, we have established the Davydov-splitting of $A_\mathrm{g}$ mode of monolayer into $N$-fold multiplet in $N$-layer \ce{CrI3} and discovered, distinct from non-magnetic few-layer atomic crystals \cite{RN23, RN24, RN25, RN26}, a unique layered-magnetism-assisted phonon scattering mechanism in the magnetic phases of \ce{CrI3}. We find this mechanism extend beyond $1^\mathrm{st}$-order phonons to the multi-phonon modes, and further resolve the distinct magnetic field dependence for different $1^\mathrm{st}$-order modes within the $N$-fold multiplet in $N$-layer \ce{CrI3}. Our calculations based on the combination of Davydov-splitting and layered-magnetism-phonon coupling successfully explain the selection rules for individual split modes and capture the rich behavior of their distinct magnetic field dependence, effective for 2D \ce{CrI3} of arbitrary thickness. \\

\noindent \textbf{Materials and Methods}\\
\noindent \textbf{Sample fabrication} \hspace{3pt} \ce{CrI3} single crystals were grown by the chemical vapor transport method, as detailed in Ref. \cite{RN18}. 1--4L \ce{CrI3} samples were exfoliated in a nitrogen-filled glovebox. Using a polymer-stamping transfer technique inside the glovebox, 1--4L \ce{CrI3} flakes were then sandwiched between two few-layer hBN flakes and transferred onto \ce{SiO2}/Si substrates for Raman spectroscopy measurements.

\noindent \textbf{Micro-Raman spectroscopy} \hspace{3pt} Micro-Raman spectroscopy measurements were carried out using a 633 nm excitation laser. The incident beam was focused by a $40\times$ objective down to $\sim$3 $\mu$m in diameter at the sample site, and the power was kept at 80 $\mu$W. The scattered light was collected by the objective in a backscattering geometry, then dispersed by a Horiba LabRAM HR Evolution Raman spectrometer, and finally detected by a thermoelectric cooled CCD camera. A closed-cycle helium cryostat is interfaced with the micro-Raman system for the temperature-dependent measurements. All thermal cycles were performed at a base pressure that is lower than $7\times10^{-7}$ mbar. In addition, a cryogen-free magnet is integrated with the low temperature cryostat for the magnetic field-dependent measurements. In this experiment, the magnetic field was applied along the out-of-plane direction and covered a range of 0 -- 2.2 Tesla.\\

\noindent \textbf{Author contributions}\\
\noindent Liuyan Zhao, Rui He, and Wencan Jin conceived the idea and designed the experiment. Zhipeng Ye, Gaihua Ye, and Laura Rojas took the experimental data under the guidance of Liuyan Zhao, Rui He, and Wencan Jin. Wencan Jin, Xiangpeng Luo, Kai Sun, and Liuyan Zhao analyzed the data and performed the calculations of phonon mode split and magnetism-phonon coupling. Bowen Yang, Fangzhou Yin, and Hyun Ho Kim fabricated thin films of \ce{CrI3} under the guidance of Adam W. Tsen. Shangjie Tian, Yang Fu, and Shaohua Yan grew bulk \ce{CrI3} single crystals. Wencan Jin, Xiangpeng Luo, Rui He, and Liuyan Zhao wrote the manuscript.\\

\vspace{20pt}

\noindent \textbf{Acknowledgements}\\
\noindent L. Zhao acknowledges support by NSF CAREER Grant No. DMR-1749774. R. He acknowledges support by NSF CAREER Grant No. DMR-1760668 and NSF MRI Grant No. DMR-1337207. A. W. Tsen acknowledges support from the US Army Research Office (W911NF-19-10267), Ontario Early Researcher Award (ER17-13-199), and the National Science and Engineering Research Council of Canada (RGPIN-2017-03815). This research was undertaken thanks in part to funding from the Canada First Research Excellence Fund. K. Sun acknowledges support through NSF Grant No. NSF-EFMA-1741618. H. Lei acknowledges support by the National Key R\&D Program of China (Grant No. 2018YFE0202600 and 2016YFA0300504), the National Natural Science Foundation of China (No. 11574394, 11774423, and 11822412), the Fundamental Research Funds for the Central Universities, and the Research Funds of Renmin University of China (18XNLG14, 19XNLG17, and 20XHN062).


\bibliographystyle{apsrev4-1}
\nocite{apsrev41Control}
\bibliography{arXiv.bib} 

\begin{thebibliography}{31}%
\makeatletter
\providecommand \@ifxundefined [1]{%
 \@ifx{#1\undefined}
}%
\providecommand \@ifnum [1]{%
 \ifnum #1\expandafter \@firstoftwo
 \else \expandafter \@secondoftwo
 \fi
}%
\providecommand \@ifx [1]{%
 \ifx #1\expandafter \@firstoftwo
 \else \expandafter \@secondoftwo
 \fi
}%
\providecommand \natexlab [1]{#1}%
\providecommand \enquote  [1]{``#1''}%
\providecommand \bibnamefont  [1]{#1}%
\providecommand \bibfnamefont [1]{#1}%
\providecommand \citenamefont [1]{#1}%
\providecommand \href@noop [0]{\@secondoftwo}%
\providecommand \href [0]{\begingroup \@sanitize@url \@href}%
\providecommand \@href[1]{\@@startlink{#1}\@@href}%
\providecommand \@@href[1]{\endgroup#1\@@endlink}%
\providecommand \@sanitize@url [0]{\catcode `\\12\catcode `\$12\catcode
  `\&12\catcode `\#12\catcode `\^12\catcode `\_12\catcode `\%12\relax}%
\providecommand \@@startlink[1]{}%
\providecommand \@@endlink[0]{}%
\providecommand \url  [0]{\begingroup\@sanitize@url \@url }%
\providecommand \@url [1]{\endgroup\@href {#1}{\urlprefix }}%
\providecommand \urlprefix  [0]{URL }%
\providecommand \Eprint [0]{\href }%
\providecommand \doibase [0]{http://dx.doi.org/}%
\providecommand \selectlanguage [0]{\@gobble}%
\providecommand \bibinfo  [0]{\@secondoftwo}%
\providecommand \bibfield  [0]{\@secondoftwo}%
\providecommand \translation [1]{[#1]}%
\providecommand \BibitemOpen [0]{}%
\providecommand \bibitemStop [0]{}%
\providecommand \bibitemNoStop [0]{.\EOS\space}%
\providecommand \EOS [0]{\spacefactor3000\relax}%
\providecommand \BibitemShut  [1]{\csname bibitem#1\endcsname}%
\let\auto@bib@innerbib\@empty
\bibitem [{\citenamefont {Klein}\ \emph {et~al.}(2018)\citenamefont {Klein},
  \citenamefont {MacNeill}, \citenamefont {Lado}, \citenamefont {Soriano},
  \citenamefont {Navarro-Moratalla}, \citenamefont {Watanabe}, \citenamefont
  {Taniguchi}, \citenamefont {Manni}, \citenamefont {Canfield}, \citenamefont
  {Fern\'{a}ndez-Rossier},\ and\ \citenamefont {Jarillo-Herrero}}]{RN1}%
  \BibitemOpen
  \bibfield  {author} {\bibinfo {author} {\bibfnamefont {D.~R.}\ \bibnamefont
  {Klein}}, \bibinfo {author} {\bibfnamefont {D.}~\bibnamefont {MacNeill}},
  \bibinfo {author} {\bibfnamefont {J.~L.}\ \bibnamefont {Lado}}, \bibinfo
  {author} {\bibfnamefont {D.}~\bibnamefont {Soriano}}, \bibinfo {author}
  {\bibfnamefont {E.}~\bibnamefont {Navarro-Moratalla}}, \bibinfo {author}
  {\bibfnamefont {K.}~\bibnamefont {Watanabe}}, \bibinfo {author}
  {\bibfnamefont {T.}~\bibnamefont {Taniguchi}}, \bibinfo {author}
  {\bibfnamefont {S.}~\bibnamefont {Manni}}, \bibinfo {author} {\bibfnamefont
  {P.}~\bibnamefont {Canfield}}, \bibinfo {author} {\bibfnamefont
  {J.}~\bibnamefont {Fern\'{a}ndez-Rossier}}, \ and\ \bibinfo {author}
  {\bibfnamefont {P.}~\bibnamefont {Jarillo-Herrero}},\ }\bibfield  {title}
  {\enquote {\bibinfo {title} {Probing magnetism in {2D} van der {Waals}
  crystalline insulators via electron tunneling},}\ }\href {\doibase
  10.1126/science.aar3617} {\bibfield  {journal} {\bibinfo  {journal}
  {Science}\ }\textbf {\bibinfo {volume} {360}},\ \bibinfo {pages} {1218--1222}
  (\bibinfo {year} {2018})}\BibitemShut {NoStop}%
\bibitem [{\citenamefont {Song}\ \emph {et~al.}(2018)\citenamefont {Song},
  \citenamefont {Cai}, \citenamefont {Tu}, \citenamefont {Zhang}, \citenamefont
  {Huang}, \citenamefont {Wilson}, \citenamefont {Seyler}, \citenamefont {Zhu},
  \citenamefont {Taniguchi}, \citenamefont {Watanabe}, \citenamefont {McGuire},
  \citenamefont {Cobden}, \citenamefont {Xiao}, \citenamefont {Yao},\ and\
  \citenamefont {Xu}}]{RN2}%
  \BibitemOpen
  \bibfield  {author} {\bibinfo {author} {\bibfnamefont {T.}~\bibnamefont
  {Song}}, \bibinfo {author} {\bibfnamefont {X.}~\bibnamefont {Cai}}, \bibinfo
  {author} {\bibfnamefont {M.~W.-Y.}\ \bibnamefont {Tu}}, \bibinfo {author}
  {\bibfnamefont {X.}~\bibnamefont {Zhang}}, \bibinfo {author} {\bibfnamefont
  {B.}~\bibnamefont {Huang}}, \bibinfo {author} {\bibfnamefont {N.~P.}\
  \bibnamefont {Wilson}}, \bibinfo {author} {\bibfnamefont {K.~L.}\
  \bibnamefont {Seyler}}, \bibinfo {author} {\bibfnamefont {L.}~\bibnamefont
  {Zhu}}, \bibinfo {author} {\bibfnamefont {T.}~\bibnamefont {Taniguchi}},
  \bibinfo {author} {\bibfnamefont {K.}~\bibnamefont {Watanabe}}, \bibinfo
  {author} {\bibfnamefont {M.~A.}\ \bibnamefont {McGuire}}, \bibinfo {author}
  {\bibfnamefont {D.~H.}\ \bibnamefont {Cobden}}, \bibinfo {author}
  {\bibfnamefont {D.}~\bibnamefont {Xiao}}, \bibinfo {author} {\bibfnamefont
  {W.}~\bibnamefont {Yao}}, \ and\ \bibinfo {author} {\bibfnamefont
  {X.}~\bibnamefont {Xu}},\ }\bibfield  {title} {\enquote {\bibinfo {title}
  {Giant tunneling magnetoresistance in spin-filter van der {Waals}
  heterostructures},}\ }\href {\doibase 10.1126/science.aar4851} {\bibfield
  {journal} {\bibinfo  {journal} {Science}\ }\textbf {\bibinfo {volume}
  {360}},\ \bibinfo {pages} {1214--1218} (\bibinfo {year} {2018})}\BibitemShut
  {NoStop}%
\bibitem [{\citenamefont {Kim}\ \emph {et~al.}(2018)\citenamefont {Kim},
  \citenamefont {Yang}, \citenamefont {Patel}, \citenamefont {Sfigakis},
  \citenamefont {Li}, \citenamefont {Tian}, \citenamefont {Lei},\ and\
  \citenamefont {Tsen}}]{RN3}%
  \BibitemOpen
  \bibfield  {author} {\bibinfo {author} {\bibfnamefont {H.~H.}\ \bibnamefont
  {Kim}}, \bibinfo {author} {\bibfnamefont {B.}~\bibnamefont {Yang}}, \bibinfo
  {author} {\bibfnamefont {T.}~\bibnamefont {Patel}}, \bibinfo {author}
  {\bibfnamefont {F.}~\bibnamefont {Sfigakis}}, \bibinfo {author}
  {\bibfnamefont {C.}~\bibnamefont {Li}}, \bibinfo {author} {\bibfnamefont
  {S.}~\bibnamefont {Tian}}, \bibinfo {author} {\bibfnamefont {H.}~\bibnamefont
  {Lei}}, \ and\ \bibinfo {author} {\bibfnamefont {A.~W.}\ \bibnamefont
  {Tsen}},\ }\bibfield  {title} {\enquote {\bibinfo {title} {One million
  percent tunnel magnetoresistance in a magnetic van der {Waals}
  heterostructure},}\ }\href {\doibase 10.1021/acs.nanolett.8b01552} {\bibfield
   {journal} {\bibinfo  {journal} {Nano Letters}\ }\textbf {\bibinfo {volume}
  {18}},\ \bibinfo {pages} {4885--4890} (\bibinfo {year} {2018})}\BibitemShut
  {NoStop}%
\bibitem [{\citenamefont {Wang}\ \emph {et~al.}(2018)\citenamefont {Wang},
  \citenamefont {Guti\'{e}rrez-Lezama}, \citenamefont {Ubrig}, \citenamefont
  {Kroner}, \citenamefont {Gibertini}, \citenamefont {Taniguchi}, \citenamefont
  {Watanabe}, \citenamefont {Imamo\u{g}lu}, \citenamefont {Giannini},\ and\
  \citenamefont {Morpurgo}}]{RN4}%
  \BibitemOpen
  \bibfield  {author} {\bibinfo {author} {\bibfnamefont {Z.}~\bibnamefont
  {Wang}}, \bibinfo {author} {\bibfnamefont {I.}~\bibnamefont
  {Guti\'{e}rrez-Lezama}}, \bibinfo {author} {\bibfnamefont {N.}~\bibnamefont
  {Ubrig}}, \bibinfo {author} {\bibfnamefont {M.}~\bibnamefont {Kroner}},
  \bibinfo {author} {\bibfnamefont {M.}~\bibnamefont {Gibertini}}, \bibinfo
  {author} {\bibfnamefont {T.}~\bibnamefont {Taniguchi}}, \bibinfo {author}
  {\bibfnamefont {K.}~\bibnamefont {Watanabe}}, \bibinfo {author}
  {\bibfnamefont {A.}~\bibnamefont {Imamo\u{g}lu}}, \bibinfo {author}
  {\bibfnamefont {E.}~\bibnamefont {Giannini}}, \ and\ \bibinfo {author}
  {\bibfnamefont {A.~F.}\ \bibnamefont {Morpurgo}},\ }\bibfield  {title}
  {\enquote {\bibinfo {title} {Very large tunneling magnetoresistance in
  layered magnetic semiconductor \ce{CrI3}},}\ }\href {\doibase
  10.1038/s41467-018-04953-8} {\bibfield  {journal} {\bibinfo  {journal}
  {Nature Communications}\ }\textbf {\bibinfo {volume} {9}},\ \bibinfo {pages}
  {2516} (\bibinfo {year} {2018})}\BibitemShut {NoStop}%
\bibitem [{\citenamefont {Huang}\ \emph {et~al.}(2017)\citenamefont {Huang},
  \citenamefont {Clark}, \citenamefont {Navarro-Moratalla}, \citenamefont
  {Klein}, \citenamefont {Cheng}, \citenamefont {Seyler}, \citenamefont
  {Zhong}, \citenamefont {Schmidgall}, \citenamefont {McGuire}, \citenamefont
  {Cobden}, \citenamefont {Yao}, \citenamefont {Xiao}, \citenamefont
  {Jarillo-Herrero},\ and\ \citenamefont {Xu}}]{RN5}%
  \BibitemOpen
  \bibfield  {author} {\bibinfo {author} {\bibfnamefont {B.}~\bibnamefont
  {Huang}}, \bibinfo {author} {\bibfnamefont {G.}~\bibnamefont {Clark}},
  \bibinfo {author} {\bibfnamefont {E.}~\bibnamefont {Navarro-Moratalla}},
  \bibinfo {author} {\bibfnamefont {D.~R.}\ \bibnamefont {Klein}}, \bibinfo
  {author} {\bibfnamefont {R.}~\bibnamefont {Cheng}}, \bibinfo {author}
  {\bibfnamefont {K.~L.}\ \bibnamefont {Seyler}}, \bibinfo {author}
  {\bibfnamefont {D.}~\bibnamefont {Zhong}}, \bibinfo {author} {\bibfnamefont
  {E.}~\bibnamefont {Schmidgall}}, \bibinfo {author} {\bibfnamefont {M.~A.}\
  \bibnamefont {McGuire}}, \bibinfo {author} {\bibfnamefont {D.~H.}\
  \bibnamefont {Cobden}}, \bibinfo {author} {\bibfnamefont {W.}~\bibnamefont
  {Yao}}, \bibinfo {author} {\bibfnamefont {D.}~\bibnamefont {Xiao}}, \bibinfo
  {author} {\bibfnamefont {P.}~\bibnamefont {Jarillo-Herrero}}, \ and\ \bibinfo
  {author} {\bibfnamefont {X.}~\bibnamefont {Xu}},\ }\bibfield  {title}
  {\enquote {\bibinfo {title} {Layer-dependent ferromagnetism in a van der
  {Waals} crystal down to the monolayer limit},}\ }\href {\doibase
  10.1038/nature22391} {\bibfield  {journal} {\bibinfo  {journal} {Nature}\
  }\textbf {\bibinfo {volume} {546}},\ \bibinfo {pages} {270--273} (\bibinfo
  {year} {2017})}\BibitemShut {NoStop}%
\bibitem [{\citenamefont {Kim}\ \emph {et~al.}(2019)\citenamefont {Kim},
  \citenamefont {Yang}, \citenamefont {Li}, \citenamefont {Jiang},
  \citenamefont {Jin}, \citenamefont {Tao}, \citenamefont {Nichols},
  \citenamefont {Sfigakis}, \citenamefont {Zhong}, \citenamefont {Li},
  \citenamefont {Tian}, \citenamefont {Cory}, \citenamefont {Miao},
  \citenamefont {Shan}, \citenamefont {Mak}, \citenamefont {Lei}, \citenamefont
  {Sun}, \citenamefont {Zhao},\ and\ \citenamefont {Tsen}}]{RN6}%
  \BibitemOpen
  \bibfield  {author} {\bibinfo {author} {\bibfnamefont {H.~H.}\ \bibnamefont
  {Kim}}, \bibinfo {author} {\bibfnamefont {B.}~\bibnamefont {Yang}}, \bibinfo
  {author} {\bibfnamefont {S.}~\bibnamefont {Li}}, \bibinfo {author}
  {\bibfnamefont {S.}~\bibnamefont {Jiang}}, \bibinfo {author} {\bibfnamefont
  {C.}~\bibnamefont {Jin}}, \bibinfo {author} {\bibfnamefont {Z.}~\bibnamefont
  {Tao}}, \bibinfo {author} {\bibfnamefont {G.}~\bibnamefont {Nichols}},
  \bibinfo {author} {\bibfnamefont {F.}~\bibnamefont {Sfigakis}}, \bibinfo
  {author} {\bibfnamefont {S.}~\bibnamefont {Zhong}}, \bibinfo {author}
  {\bibfnamefont {C.}~\bibnamefont {Li}}, \bibinfo {author} {\bibfnamefont
  {S.}~\bibnamefont {Tian}}, \bibinfo {author} {\bibfnamefont {D.~G.}\
  \bibnamefont {Cory}}, \bibinfo {author} {\bibfnamefont {G.-X.}\ \bibnamefont
  {Miao}}, \bibinfo {author} {\bibfnamefont {J.}~\bibnamefont {Shan}}, \bibinfo
  {author} {\bibfnamefont {K.~F.}\ \bibnamefont {Mak}}, \bibinfo {author}
  {\bibfnamefont {H.}~\bibnamefont {Lei}}, \bibinfo {author} {\bibfnamefont
  {K.}~\bibnamefont {Sun}}, \bibinfo {author} {\bibfnamefont {L.}~\bibnamefont
  {Zhao}}, \ and\ \bibinfo {author} {\bibfnamefont {A.~W.}\ \bibnamefont
  {Tsen}},\ }\bibfield  {title} {\enquote {\bibinfo {title} {Evolution of
  interlayer and intralayer magnetism in three atomically thin chromium
  trihalides},}\ }\href {\doibase 10.1073/pnas.1902100116} {\bibfield
  {journal} {\bibinfo  {journal} {Proceedings of the National Academy of
  Sciences}\ }\textbf {\bibinfo {volume} {116}},\ \bibinfo {pages} {11131}
  (\bibinfo {year} {2019})}\BibitemShut {NoStop}%
\bibitem [{\citenamefont {Li}\ \emph {et~al.}(2020)\citenamefont {Li},
  \citenamefont {Ye}, \citenamefont {Luo}, \citenamefont {Ye}, \citenamefont
  {Kim}, \citenamefont {Yang}, \citenamefont {Tian}, \citenamefont {Li},
  \citenamefont {Lei}, \citenamefont {Tsen}, \citenamefont {Sun}, \citenamefont
  {He},\ and\ \citenamefont {Zhao}}]{RN7}%
  \BibitemOpen
  \bibfield  {author} {\bibinfo {author} {\bibfnamefont {S.}~\bibnamefont
  {Li}}, \bibinfo {author} {\bibfnamefont {Z.}~\bibnamefont {Ye}}, \bibinfo
  {author} {\bibfnamefont {X.}~\bibnamefont {Luo}}, \bibinfo {author}
  {\bibfnamefont {G.}~\bibnamefont {Ye}}, \bibinfo {author} {\bibfnamefont
  {H.~H.}\ \bibnamefont {Kim}}, \bibinfo {author} {\bibfnamefont
  {B.}~\bibnamefont {Yang}}, \bibinfo {author} {\bibfnamefont {S.}~\bibnamefont
  {Tian}}, \bibinfo {author} {\bibfnamefont {C.}~\bibnamefont {Li}}, \bibinfo
  {author} {\bibfnamefont {H.}~\bibnamefont {Lei}}, \bibinfo {author}
  {\bibfnamefont {A.~W.}\ \bibnamefont {Tsen}}, \bibinfo {author}
  {\bibfnamefont {K.}~\bibnamefont {Sun}}, \bibinfo {author} {\bibfnamefont
  {R.}~\bibnamefont {He}}, \ and\ \bibinfo {author} {\bibfnamefont
  {L.}~\bibnamefont {Zhao}},\ }\bibfield  {title} {\enquote {\bibinfo {title}
  {Magnetic-field-induced quantum phase transitions in a van der {Waals}
  magnet},}\ }\href {\doibase 10.1103/PhysRevX.10.011075} {\bibfield  {journal}
  {\bibinfo  {journal} {Phys. Rev. X}\ }\textbf {\bibinfo {volume} {10}},\
  \bibinfo {pages} {011075} (\bibinfo {year} {2020})}\BibitemShut {NoStop}%
\bibitem [{\citenamefont {Huang}\ \emph {et~al.}(2018)\citenamefont {Huang},
  \citenamefont {Clark}, \citenamefont {Klein}, \citenamefont {MacNeill},
  \citenamefont {Navarro-Moratalla}, \citenamefont {Seyler}, \citenamefont
  {Wilson}, \citenamefont {McGuire}, \citenamefont {Cobden}, \citenamefont
  {Xiao}, \citenamefont {Yao}, \citenamefont {Jarillo-Herrero},\ and\
  \citenamefont {Xu}}]{RN8}%
  \BibitemOpen
  \bibfield  {author} {\bibinfo {author} {\bibfnamefont {B.}~\bibnamefont
  {Huang}}, \bibinfo {author} {\bibfnamefont {G.}~\bibnamefont {Clark}},
  \bibinfo {author} {\bibfnamefont {D.~R.}\ \bibnamefont {Klein}}, \bibinfo
  {author} {\bibfnamefont {D.}~\bibnamefont {MacNeill}}, \bibinfo {author}
  {\bibfnamefont {E.}~\bibnamefont {Navarro-Moratalla}}, \bibinfo {author}
  {\bibfnamefont {K.~L.}\ \bibnamefont {Seyler}}, \bibinfo {author}
  {\bibfnamefont {N.}~\bibnamefont {Wilson}}, \bibinfo {author} {\bibfnamefont
  {M.~A.}\ \bibnamefont {McGuire}}, \bibinfo {author} {\bibfnamefont {D.~H.}\
  \bibnamefont {Cobden}}, \bibinfo {author} {\bibfnamefont {D.}~\bibnamefont
  {Xiao}}, \bibinfo {author} {\bibfnamefont {W.}~\bibnamefont {Yao}}, \bibinfo
  {author} {\bibfnamefont {P.}~\bibnamefont {Jarillo-Herrero}}, \ and\ \bibinfo
  {author} {\bibfnamefont {X.}~\bibnamefont {Xu}},\ }\bibfield  {title}
  {\enquote {\bibinfo {title} {Electrical control of {2D} magnetism in bilayer
  \ce{CrI3}},}\ }\href {\doibase 10.1038/s41565-018-0121-3} {\bibfield
  {journal} {\bibinfo  {journal} {Nature Nanotechnology}\ }\textbf {\bibinfo
  {volume} {13}},\ \bibinfo {pages} {544--548} (\bibinfo {year}
  {2018})}\BibitemShut {NoStop}%
\bibitem [{\citenamefont {Jiang}\ \emph
  {et~al.}(2018{\natexlab{a}})\citenamefont {Jiang}, \citenamefont {Shan},\
  and\ \citenamefont {Mak}}]{RN9}%
  \BibitemOpen
  \bibfield  {author} {\bibinfo {author} {\bibfnamefont {S.}~\bibnamefont
  {Jiang}}, \bibinfo {author} {\bibfnamefont {J.}~\bibnamefont {Shan}}, \ and\
  \bibinfo {author} {\bibfnamefont {K.~F.}\ \bibnamefont {Mak}},\ }\bibfield
  {title} {\enquote {\bibinfo {title} {Electric-field switching of
  two-dimensional van der {Waals} magnets},}\ }\href {\doibase
  10.1038/s41563-018-0040-6} {\bibfield  {journal} {\bibinfo  {journal} {Nature
  Materials}\ }\textbf {\bibinfo {volume} {17}},\ \bibinfo {pages} {406--410}
  (\bibinfo {year} {2018}{\natexlab{a}})}\BibitemShut {NoStop}%
\bibitem [{\citenamefont {Kim}\ \emph {et~al.}(2020)\citenamefont {Kim},
  \citenamefont {Jiang}, \citenamefont {Yang}, \citenamefont {Zhong},
  \citenamefont {Tian}, \citenamefont {Li}, \citenamefont {Lei}, \citenamefont
  {Shan}, \citenamefont {Mak},\ and\ \citenamefont {Tsen}}]{RN10A}%
  \BibitemOpen
  \bibfield  {author} {\bibinfo {author} {\bibfnamefont {H.~H.}\ \bibnamefont
  {Kim}}, \bibinfo {author} {\bibfnamefont {S.}~\bibnamefont {Jiang}}, \bibinfo
  {author} {\bibfnamefont {B.}~\bibnamefont {Yang}}, \bibinfo {author}
  {\bibfnamefont {S.}~\bibnamefont {Zhong}}, \bibinfo {author} {\bibfnamefont
  {S.}~\bibnamefont {Tian}}, \bibinfo {author} {\bibfnamefont {C.}~\bibnamefont
  {Li}}, \bibinfo {author} {\bibfnamefont {H.}~\bibnamefont {Lei}}, \bibinfo
  {author} {\bibfnamefont {J.}~\bibnamefont {Shan}}, \bibinfo {author}
  {\bibfnamefont {K.~F.}\ \bibnamefont {Mak}}, \ and\ \bibinfo {author}
  {\bibfnamefont {A.~W.}\ \bibnamefont {Tsen}},\ }\bibfield  {title} {\enquote
  {\bibinfo {title} {Magneto-memristive switching in a {2D} layer
  antiferromagnet},}\ }\href {\doibase 10.1002/adma.201905433} {\bibfield
  {journal} {\bibinfo  {journal} {Advanced Materials}\ }\textbf {\bibinfo
  {volume} {32}},\ \bibinfo {pages} {1905433} (\bibinfo {year}
  {2020})}\BibitemShut {NoStop}%
\bibitem [{\citenamefont {Jiang}\ \emph
  {et~al.}(2018{\natexlab{b}})\citenamefont {Jiang}, \citenamefont {Li},
  \citenamefont {Wang}, \citenamefont {Mak},\ and\ \citenamefont
  {Shan}}]{RN10}%
  \BibitemOpen
  \bibfield  {author} {\bibinfo {author} {\bibfnamefont {S.}~\bibnamefont
  {Jiang}}, \bibinfo {author} {\bibfnamefont {L.}~\bibnamefont {Li}}, \bibinfo
  {author} {\bibfnamefont {Z.}~\bibnamefont {Wang}}, \bibinfo {author}
  {\bibfnamefont {K.~F.}\ \bibnamefont {Mak}}, \ and\ \bibinfo {author}
  {\bibfnamefont {J.}~\bibnamefont {Shan}},\ }\bibfield  {title} {\enquote
  {\bibinfo {title} {Controlling magnetism in {2D} \ce{CrI3} by electrostatic
  doping},}\ }\href {\doibase 10.1038/s41565-018-0135-x} {\bibfield  {journal}
  {\bibinfo  {journal} {Nature Nanotechnology}\ }\textbf {\bibinfo {volume}
  {13}},\ \bibinfo {pages} {549--553} (\bibinfo {year}
  {2018}{\natexlab{b}})}\BibitemShut {NoStop}%
\bibitem [{\citenamefont {Song}\ \emph {et~al.}(2019)\citenamefont {Song},
  \citenamefont {Fei}, \citenamefont {Yankowitz}, \citenamefont {Lin},
  \citenamefont {Jiang}, \citenamefont {Hwangbo}, \citenamefont {Zhang},
  \citenamefont {Sun}, \citenamefont {Taniguchi}, \citenamefont {Watanabe},
  \citenamefont {McGuire}, \citenamefont {Graf}, \citenamefont {Cao},
  \citenamefont {Chu}, \citenamefont {Cobden}, \citenamefont {Dean},
  \citenamefont {Xiao},\ and\ \citenamefont {Xu}}]{RN11}%
  \BibitemOpen
  \bibfield  {author} {\bibinfo {author} {\bibfnamefont {T.}~\bibnamefont
  {Song}}, \bibinfo {author} {\bibfnamefont {Z.}~\bibnamefont {Fei}}, \bibinfo
  {author} {\bibfnamefont {M.}~\bibnamefont {Yankowitz}}, \bibinfo {author}
  {\bibfnamefont {Z.}~\bibnamefont {Lin}}, \bibinfo {author} {\bibfnamefont
  {Q.}~\bibnamefont {Jiang}}, \bibinfo {author} {\bibfnamefont
  {K.}~\bibnamefont {Hwangbo}}, \bibinfo {author} {\bibfnamefont
  {Q.}~\bibnamefont {Zhang}}, \bibinfo {author} {\bibfnamefont
  {B.}~\bibnamefont {Sun}}, \bibinfo {author} {\bibfnamefont {T.}~\bibnamefont
  {Taniguchi}}, \bibinfo {author} {\bibfnamefont {K.}~\bibnamefont {Watanabe}},
  \bibinfo {author} {\bibfnamefont {M.~A.}\ \bibnamefont {McGuire}}, \bibinfo
  {author} {\bibfnamefont {D.}~\bibnamefont {Graf}}, \bibinfo {author}
  {\bibfnamefont {T.}~\bibnamefont {Cao}}, \bibinfo {author} {\bibfnamefont
  {J.-H.}\ \bibnamefont {Chu}}, \bibinfo {author} {\bibfnamefont {D.~H.}\
  \bibnamefont {Cobden}}, \bibinfo {author} {\bibfnamefont {C.~R.}\
  \bibnamefont {Dean}}, \bibinfo {author} {\bibfnamefont {D.}~\bibnamefont
  {Xiao}}, \ and\ \bibinfo {author} {\bibfnamefont {X.}~\bibnamefont {Xu}},\
  }\bibfield  {title} {\enquote {\bibinfo {title} {Switching {2D} magnetic
  states via pressure tuning of layer stacking},}\ }\href {\doibase
  10.1038/s41563-019-0505-2} {\bibfield  {journal} {\bibinfo  {journal} {Nature
  Materials}\ }\textbf {\bibinfo {volume} {18}},\ \bibinfo {pages} {1298--1302}
  (\bibinfo {year} {2019})}\BibitemShut {NoStop}%
\bibitem [{\citenamefont {Li}\ \emph {et~al.}(2019)\citenamefont {Li},
  \citenamefont {Jiang}, \citenamefont {Sivadas}, \citenamefont {Wang},
  \citenamefont {Xu}, \citenamefont {Weber}, \citenamefont {Goldberger},
  \citenamefont {Watanabe}, \citenamefont {Taniguchi}, \citenamefont {Fennie},
  \citenamefont {Fai~Mak},\ and\ \citenamefont {Shan}}]{RN12}%
  \BibitemOpen
  \bibfield  {author} {\bibinfo {author} {\bibfnamefont {T.}~\bibnamefont
  {Li}}, \bibinfo {author} {\bibfnamefont {S.}~\bibnamefont {Jiang}}, \bibinfo
  {author} {\bibfnamefont {N.}~\bibnamefont {Sivadas}}, \bibinfo {author}
  {\bibfnamefont {Z.}~\bibnamefont {Wang}}, \bibinfo {author} {\bibfnamefont
  {Y.}~\bibnamefont {Xu}}, \bibinfo {author} {\bibfnamefont {D.}~\bibnamefont
  {Weber}}, \bibinfo {author} {\bibfnamefont {J.~E.}\ \bibnamefont
  {Goldberger}}, \bibinfo {author} {\bibfnamefont {K.}~\bibnamefont
  {Watanabe}}, \bibinfo {author} {\bibfnamefont {T.}~\bibnamefont {Taniguchi}},
  \bibinfo {author} {\bibfnamefont {C.~J.}\ \bibnamefont {Fennie}}, \bibinfo
  {author} {\bibfnamefont {K.}~\bibnamefont {Fai~Mak}}, \ and\ \bibinfo
  {author} {\bibfnamefont {J.}~\bibnamefont {Shan}},\ }\bibfield  {title}
  {\enquote {\bibinfo {title} {Pressure-controlled interlayer magnetism in
  atomically thin \ce{CrI3}},}\ }\href {\doibase 10.1038/s41563-019-0506-1}
  {\bibfield  {journal} {\bibinfo  {journal} {Nature Materials}\ }\textbf
  {\bibinfo {volume} {18}},\ \bibinfo {pages} {1303--1308} (\bibinfo {year}
  {2019})}\BibitemShut {NoStop}%
\bibitem [{\citenamefont {Seyler}\ \emph {et~al.}(2018)\citenamefont {Seyler},
  \citenamefont {Zhong}, \citenamefont {Klein}, \citenamefont {Gao},
  \citenamefont {Zhang}, \citenamefont {Huang}, \citenamefont
  {Navarro-Moratalla}, \citenamefont {Yang}, \citenamefont {Cobden},
  \citenamefont {McGuire}, \citenamefont {Yao}, \citenamefont {Xiao},
  \citenamefont {Jarillo-Herrero},\ and\ \citenamefont {Xu}}]{RN13}%
  \BibitemOpen
  \bibfield  {author} {\bibinfo {author} {\bibfnamefont {K.~L.}\ \bibnamefont
  {Seyler}}, \bibinfo {author} {\bibfnamefont {D.}~\bibnamefont {Zhong}},
  \bibinfo {author} {\bibfnamefont {D.~R.}\ \bibnamefont {Klein}}, \bibinfo
  {author} {\bibfnamefont {S.}~\bibnamefont {Gao}}, \bibinfo {author}
  {\bibfnamefont {X.}~\bibnamefont {Zhang}}, \bibinfo {author} {\bibfnamefont
  {B.}~\bibnamefont {Huang}}, \bibinfo {author} {\bibfnamefont
  {E.}~\bibnamefont {Navarro-Moratalla}}, \bibinfo {author} {\bibfnamefont
  {L.}~\bibnamefont {Yang}}, \bibinfo {author} {\bibfnamefont {D.~H.}\
  \bibnamefont {Cobden}}, \bibinfo {author} {\bibfnamefont {M.~A.}\
  \bibnamefont {McGuire}}, \bibinfo {author} {\bibfnamefont {W.}~\bibnamefont
  {Yao}}, \bibinfo {author} {\bibfnamefont {D.}~\bibnamefont {Xiao}}, \bibinfo
  {author} {\bibfnamefont {P.}~\bibnamefont {Jarillo-Herrero}}, \ and\ \bibinfo
  {author} {\bibfnamefont {X.}~\bibnamefont {Xu}},\ }\bibfield  {title}
  {\enquote {\bibinfo {title} {Ligand-field helical luminescence in a {2D}
  ferromagnetic insulator},}\ }\href {\doibase 10.1038/s41567-017-0006-7}
  {\bibfield  {journal} {\bibinfo  {journal} {Nature Physics}\ }\textbf
  {\bibinfo {volume} {14}},\ \bibinfo {pages} {277--281} (\bibinfo {year}
  {2018})}\BibitemShut {NoStop}%
\bibitem [{\citenamefont {Sun}\ \emph {et~al.}(2019)\citenamefont {Sun},
  \citenamefont {Yi}, \citenamefont {Song}, \citenamefont {Clark},
  \citenamefont {Huang}, \citenamefont {Shan}, \citenamefont {Wu},
  \citenamefont {Huang}, \citenamefont {Gao}, \citenamefont {Chen},
  \citenamefont {McGuire}, \citenamefont {Cao}, \citenamefont {Xiao},
  \citenamefont {Liu}, \citenamefont {Yao}, \citenamefont {Xu},\ and\
  \citenamefont {Wu}}]{RN14}%
  \BibitemOpen
  \bibfield  {author} {\bibinfo {author} {\bibfnamefont {Z.}~\bibnamefont
  {Sun}}, \bibinfo {author} {\bibfnamefont {Y.}~\bibnamefont {Yi}}, \bibinfo
  {author} {\bibfnamefont {T.}~\bibnamefont {Song}}, \bibinfo {author}
  {\bibfnamefont {G.}~\bibnamefont {Clark}}, \bibinfo {author} {\bibfnamefont
  {B.}~\bibnamefont {Huang}}, \bibinfo {author} {\bibfnamefont
  {Y.}~\bibnamefont {Shan}}, \bibinfo {author} {\bibfnamefont {S.}~\bibnamefont
  {Wu}}, \bibinfo {author} {\bibfnamefont {D.}~\bibnamefont {Huang}}, \bibinfo
  {author} {\bibfnamefont {C.}~\bibnamefont {Gao}}, \bibinfo {author}
  {\bibfnamefont {Z.}~\bibnamefont {Chen}}, \bibinfo {author} {\bibfnamefont
  {M.}~\bibnamefont {McGuire}}, \bibinfo {author} {\bibfnamefont
  {T.}~\bibnamefont {Cao}}, \bibinfo {author} {\bibfnamefont {D.}~\bibnamefont
  {Xiao}}, \bibinfo {author} {\bibfnamefont {W.-T.}\ \bibnamefont {Liu}},
  \bibinfo {author} {\bibfnamefont {W.}~\bibnamefont {Yao}}, \bibinfo {author}
  {\bibfnamefont {X.}~\bibnamefont {Xu}}, \ and\ \bibinfo {author}
  {\bibfnamefont {S.}~\bibnamefont {Wu}},\ }\bibfield  {title} {\enquote
  {\bibinfo {title} {Giant nonreciprocal second-harmonic generation from
  antiferromagnetic bilayer \ce{CrI3}},}\ }\href {\doibase
  10.1038/s41586-019-1445-3} {\bibfield  {journal} {\bibinfo  {journal}
  {Nature}\ }\textbf {\bibinfo {volume} {572}},\ \bibinfo {pages} {497--501}
  (\bibinfo {year} {2019})}\BibitemShut {NoStop}%
\bibitem [{\citenamefont {Huang}\ \emph {et~al.}(2020)\citenamefont {Huang},
  \citenamefont {Cenker}, \citenamefont {Zhang}, \citenamefont {Ray},
  \citenamefont {Song}, \citenamefont {Taniguchi}, \citenamefont {Watanabe},
  \citenamefont {McGuire}, \citenamefont {Xiao},\ and\ \citenamefont
  {Xu}}]{RN15}%
  \BibitemOpen
  \bibfield  {author} {\bibinfo {author} {\bibfnamefont {B.}~\bibnamefont
  {Huang}}, \bibinfo {author} {\bibfnamefont {J.}~\bibnamefont {Cenker}},
  \bibinfo {author} {\bibfnamefont {X.}~\bibnamefont {Zhang}}, \bibinfo
  {author} {\bibfnamefont {E.~L.}\ \bibnamefont {Ray}}, \bibinfo {author}
  {\bibfnamefont {T.}~\bibnamefont {Song}}, \bibinfo {author} {\bibfnamefont
  {T.}~\bibnamefont {Taniguchi}}, \bibinfo {author} {\bibfnamefont
  {K.}~\bibnamefont {Watanabe}}, \bibinfo {author} {\bibfnamefont {M.~A.}\
  \bibnamefont {McGuire}}, \bibinfo {author} {\bibfnamefont {D.}~\bibnamefont
  {Xiao}}, \ and\ \bibinfo {author} {\bibfnamefont {X.}~\bibnamefont {Xu}},\
  }\bibfield  {title} {\enquote {\bibinfo {title} {Tuning inelastic light
  scattering via symmetry control in the two-dimensional magnet \ce{CrI3}},}\
  }\href {\doibase 10.1038/s41565-019-0598-4} {\bibfield  {journal} {\bibinfo
  {journal} {Nature Nanotechnology}\ }\textbf {\bibinfo {volume} {15}},\
  \bibinfo {pages} {212--216} (\bibinfo {year} {2020})}\BibitemShut {NoStop}%
\bibitem [{\citenamefont {McCreary}\ \emph {et~al.}(2020)\citenamefont
  {McCreary}, \citenamefont {Mai}, \citenamefont {Utermohlen}, \citenamefont
  {Simpson}, \citenamefont {Garrity}, \citenamefont {Feng}, \citenamefont
  {Shcherbakov}, \citenamefont {Zhu}, \citenamefont {Hu}, \citenamefont
  {Weber}, \citenamefont {Watanabe}, \citenamefont {Taniguchi}, \citenamefont
  {Goldberger}, \citenamefont {Mao}, \citenamefont {Lau}, \citenamefont {Lu},
  \citenamefont {Trivedi}, \citenamefont {Vald{\'e}s~Aguilar},\ and\
  \citenamefont {Hight~Walker}}]{RN16}%
  \BibitemOpen
  \bibfield  {author} {\bibinfo {author} {\bibfnamefont {A.}~\bibnamefont
  {McCreary}}, \bibinfo {author} {\bibfnamefont {T.~T.}\ \bibnamefont {Mai}},
  \bibinfo {author} {\bibfnamefont {F.~G.}\ \bibnamefont {Utermohlen}},
  \bibinfo {author} {\bibfnamefont {J.~R.}\ \bibnamefont {Simpson}}, \bibinfo
  {author} {\bibfnamefont {K.~F.}\ \bibnamefont {Garrity}}, \bibinfo {author}
  {\bibfnamefont {X.}~\bibnamefont {Feng}}, \bibinfo {author} {\bibfnamefont
  {D.}~\bibnamefont {Shcherbakov}}, \bibinfo {author} {\bibfnamefont
  {Y.}~\bibnamefont {Zhu}}, \bibinfo {author} {\bibfnamefont {J.}~\bibnamefont
  {Hu}}, \bibinfo {author} {\bibfnamefont {D.}~\bibnamefont {Weber}}, \bibinfo
  {author} {\bibfnamefont {K.}~\bibnamefont {Watanabe}}, \bibinfo {author}
  {\bibfnamefont {T.}~\bibnamefont {Taniguchi}}, \bibinfo {author}
  {\bibfnamefont {J.~E.}\ \bibnamefont {Goldberger}}, \bibinfo {author}
  {\bibfnamefont {Z.}~\bibnamefont {Mao}}, \bibinfo {author} {\bibfnamefont
  {C.~N.}\ \bibnamefont {Lau}}, \bibinfo {author} {\bibfnamefont
  {Y.}~\bibnamefont {Lu}}, \bibinfo {author} {\bibfnamefont {N.}~\bibnamefont
  {Trivedi}}, \bibinfo {author} {\bibfnamefont {R.}~\bibnamefont
  {Vald{\'e}s~Aguilar}}, \ and\ \bibinfo {author} {\bibfnamefont {A.~R.}\
  \bibnamefont {Hight~Walker}},\ }\bibfield  {title} {\enquote {\bibinfo
  {title} {Distinct magneto-{Raman} signatures of spin-flip phase transitions
  in \ce{CrI3}},}\ }\href {\doibase 10.1038/s41467-020-17320-3} {\bibfield
  {journal} {\bibinfo  {journal} {Nature Communications}\ }\textbf {\bibinfo
  {volume} {11}},\ \bibinfo {pages} {3879} (\bibinfo {year}
  {2020})}\BibitemShut {NoStop}%
\bibitem [{\citenamefont {Zhang}\ \emph {et~al.}(2020)\citenamefont {Zhang},
  \citenamefont {Wu}, \citenamefont {Lyu}, \citenamefont {Wu}, \citenamefont
  {Zhao}, \citenamefont {Chen}, \citenamefont {Jia}, \citenamefont {Zhang},
  \citenamefont {Wang}, \citenamefont {Wang}, \citenamefont {Chen},
  \citenamefont {Mei}, \citenamefont {Taniguchi}, \citenamefont {Watanabe},
  \citenamefont {Yan}, \citenamefont {Liu}, \citenamefont {Huang},
  \citenamefont {Zhao},\ and\ \citenamefont {Huang}}]{RN17}%
  \BibitemOpen
  \bibfield  {author} {\bibinfo {author} {\bibfnamefont {Y.}~\bibnamefont
  {Zhang}}, \bibinfo {author} {\bibfnamefont {X.}~\bibnamefont {Wu}}, \bibinfo
  {author} {\bibfnamefont {B.}~\bibnamefont {Lyu}}, \bibinfo {author}
  {\bibfnamefont {M.}~\bibnamefont {Wu}}, \bibinfo {author} {\bibfnamefont
  {S.}~\bibnamefont {Zhao}}, \bibinfo {author} {\bibfnamefont {J.}~\bibnamefont
  {Chen}}, \bibinfo {author} {\bibfnamefont {M.}~\bibnamefont {Jia}}, \bibinfo
  {author} {\bibfnamefont {C.}~\bibnamefont {Zhang}}, \bibinfo {author}
  {\bibfnamefont {L.}~\bibnamefont {Wang}}, \bibinfo {author} {\bibfnamefont
  {X.}~\bibnamefont {Wang}}, \bibinfo {author} {\bibfnamefont {Y.}~\bibnamefont
  {Chen}}, \bibinfo {author} {\bibfnamefont {J.}~\bibnamefont {Mei}}, \bibinfo
  {author} {\bibfnamefont {T.}~\bibnamefont {Taniguchi}}, \bibinfo {author}
  {\bibfnamefont {K.}~\bibnamefont {Watanabe}}, \bibinfo {author}
  {\bibfnamefont {H.}~\bibnamefont {Yan}}, \bibinfo {author} {\bibfnamefont
  {Q.}~\bibnamefont {Liu}}, \bibinfo {author} {\bibfnamefont {L.}~\bibnamefont
  {Huang}}, \bibinfo {author} {\bibfnamefont {Y.}~\bibnamefont {Zhao}}, \ and\
  \bibinfo {author} {\bibfnamefont {M.}~\bibnamefont {Huang}},\ }\bibfield
  {title} {\enquote {\bibinfo {title} {Magnetic order-induced polarization
  anomaly of {Raman} scattering in {2D} magnet \ce{CrI3}},}\ }\href {\doibase
  10.1021/acs.nanolett.9b04634} {\bibfield  {journal} {\bibinfo  {journal}
  {Nano Letters}\ }\textbf {\bibinfo {volume} {20}},\ \bibinfo {pages}
  {729--734} (\bibinfo {year} {2020})}\BibitemShut {NoStop}%
\bibitem [{\citenamefont {Jin}\ \emph {et~al.}(2018)\citenamefont {Jin},
  \citenamefont {Kim}, \citenamefont {Ye}, \citenamefont {Li}, \citenamefont
  {Rezaie}, \citenamefont {Diaz}, \citenamefont {Siddiq}, \citenamefont
  {Wauer}, \citenamefont {Yang}, \citenamefont {Li}, \citenamefont {Tian},
  \citenamefont {Sun}, \citenamefont {Lei}, \citenamefont {Tsen}, \citenamefont
  {Zhao},\ and\ \citenamefont {He}}]{RN18}%
  \BibitemOpen
  \bibfield  {author} {\bibinfo {author} {\bibfnamefont {W.}~\bibnamefont
  {Jin}}, \bibinfo {author} {\bibfnamefont {H.~H.}\ \bibnamefont {Kim}},
  \bibinfo {author} {\bibfnamefont {Z.}~\bibnamefont {Ye}}, \bibinfo {author}
  {\bibfnamefont {S.}~\bibnamefont {Li}}, \bibinfo {author} {\bibfnamefont
  {P.}~\bibnamefont {Rezaie}}, \bibinfo {author} {\bibfnamefont
  {F.}~\bibnamefont {Diaz}}, \bibinfo {author} {\bibfnamefont {S.}~\bibnamefont
  {Siddiq}}, \bibinfo {author} {\bibfnamefont {E.}~\bibnamefont {Wauer}},
  \bibinfo {author} {\bibfnamefont {B.}~\bibnamefont {Yang}}, \bibinfo {author}
  {\bibfnamefont {C.}~\bibnamefont {Li}}, \bibinfo {author} {\bibfnamefont
  {S.}~\bibnamefont {Tian}}, \bibinfo {author} {\bibfnamefont {K.}~\bibnamefont
  {Sun}}, \bibinfo {author} {\bibfnamefont {H.}~\bibnamefont {Lei}}, \bibinfo
  {author} {\bibfnamefont {A.~W.}\ \bibnamefont {Tsen}}, \bibinfo {author}
  {\bibfnamefont {L.}~\bibnamefont {Zhao}}, \ and\ \bibinfo {author}
  {\bibfnamefont {R.}~\bibnamefont {He}},\ }\bibfield  {title} {\enquote
  {\bibinfo {title} {Raman fingerprint of two terahertz spin wave branches in a
  two-dimensional honeycomb {Ising} ferromagnet},}\ }\href {\doibase
  10.1038/s41467-018-07547-6} {\bibfield  {journal} {\bibinfo  {journal}
  {Nature Communications}\ }\textbf {\bibinfo {volume} {9}},\ \bibinfo {pages}
  {5122} (\bibinfo {year} {2018})}\BibitemShut {NoStop}%
\bibitem [{\citenamefont {Larson}\ and\ \citenamefont {Kaxiras}(2018)}]{RN19}%
  \BibitemOpen
  \bibfield  {author} {\bibinfo {author} {\bibfnamefont {D.~T.}\ \bibnamefont
  {Larson}}\ and\ \bibinfo {author} {\bibfnamefont {E.}~\bibnamefont
  {Kaxiras}},\ }\bibfield  {title} {\enquote {\bibinfo {title} {Raman spectrum
  of \ce{CrI3}: {An} \textit{ab initio} study},}\ }\href
  {https://link.aps.org/doi/10.1103/PhysRevB.98.085406} {\bibfield  {journal}
  {\bibinfo  {journal} {Physical Review B}\ }\textbf {\bibinfo {volume} {98}},\
  \bibinfo {pages} {085406} (\bibinfo {year} {2018})}\BibitemShut {NoStop}%
\bibitem [{\citenamefont {Webster}\ \emph {et~al.}(2018)\citenamefont
  {Webster}, \citenamefont {Liang},\ and\ \citenamefont {Yan}}]{RN20}%
  \BibitemOpen
  \bibfield  {author} {\bibinfo {author} {\bibfnamefont {L.}~\bibnamefont
  {Webster}}, \bibinfo {author} {\bibfnamefont {L.}~\bibnamefont {Liang}}, \
  and\ \bibinfo {author} {\bibfnamefont {J.-A.}\ \bibnamefont {Yan}},\
  }\bibfield  {title} {\enquote {\bibinfo {title} {Distinct spin-lattice and
  spin-phonon interactions in monolayer magnetic \ce{CrI3}},}\ }\href {\doibase
  10.1039/C8CP03599G} {\bibfield  {journal} {\bibinfo  {journal} {Physical
  Chemistry Chemical Physics}\ }\textbf {\bibinfo {volume} {20}},\ \bibinfo
  {pages} {23546--23555} (\bibinfo {year} {2018})}\BibitemShut {NoStop}%
\bibitem [{\citenamefont {Djurdji\'{c}-Mijin}\ \emph
  {et~al.}(2018)\citenamefont {Djurdji\'{c}-Mijin}, \citenamefont
  {\u{S}olaji\'{c}}, \citenamefont {Pe\u{s}i\'{c}}, \citenamefont
  {\u{S}\'{c}epanovi\'{c}}, \citenamefont {Liu}, \citenamefont {Baum},
  \citenamefont {Petrovic}, \citenamefont {Lazarevi\'{c}},\ and\ \citenamefont
  {Popovi\'{c}}}]{RN21}%
  \BibitemOpen
  \bibfield  {author} {\bibinfo {author} {\bibfnamefont {S.}~\bibnamefont
  {Djurdji\'{c}-Mijin}}, \bibinfo {author} {\bibfnamefont {A.}~\bibnamefont
  {\u{S}olaji\'{c}}}, \bibinfo {author} {\bibfnamefont {J.}~\bibnamefont
  {Pe\u{s}i\'{c}}}, \bibinfo {author} {\bibfnamefont {M.}~\bibnamefont
  {\u{S}\'{c}epanovi\'{c}}}, \bibinfo {author} {\bibfnamefont {Y.}~\bibnamefont
  {Liu}}, \bibinfo {author} {\bibfnamefont {A.}~\bibnamefont {Baum}}, \bibinfo
  {author} {\bibfnamefont {C.}~\bibnamefont {Petrovic}}, \bibinfo {author}
  {\bibfnamefont {N.}~\bibnamefont {Lazarevi\'{c}}}, \ and\ \bibinfo {author}
  {\bibfnamefont {Z.}~\bibnamefont {Popovi\'{c}}},\ }\bibfield  {title}
  {\enquote {\bibinfo {title} {Lattice dynamics and phase transition in
  \ce{CrI3} single crystals},}\ }\href@noop {} {\bibfield  {journal} {\bibinfo
  {journal} {Physical Review B}\ }\textbf {\bibinfo {volume} {98}},\ \bibinfo
  {pages} {104307} (\bibinfo {year} {2018})}\BibitemShut {NoStop}%
\bibitem [{\citenamefont {Shcherbakov}\ \emph {et~al.}(2018)\citenamefont
  {Shcherbakov}, \citenamefont {Stepanov}, \citenamefont {Weber}, \citenamefont
  {Wang}, \citenamefont {Hu}, \citenamefont {Zhu}, \citenamefont {Watanabe},
  \citenamefont {Taniguchi}, \citenamefont {Mao}, \citenamefont {Windl},
  \citenamefont {Goldberger}, \citenamefont {Bockrath},\ and\ \citenamefont
  {Lau}}]{RN22}%
  \BibitemOpen
  \bibfield  {author} {\bibinfo {author} {\bibfnamefont {D.}~\bibnamefont
  {Shcherbakov}}, \bibinfo {author} {\bibfnamefont {P.}~\bibnamefont
  {Stepanov}}, \bibinfo {author} {\bibfnamefont {D.}~\bibnamefont {Weber}},
  \bibinfo {author} {\bibfnamefont {Y.}~\bibnamefont {Wang}}, \bibinfo {author}
  {\bibfnamefont {J.}~\bibnamefont {Hu}}, \bibinfo {author} {\bibfnamefont
  {Y.}~\bibnamefont {Zhu}}, \bibinfo {author} {\bibfnamefont {K.}~\bibnamefont
  {Watanabe}}, \bibinfo {author} {\bibfnamefont {T.}~\bibnamefont {Taniguchi}},
  \bibinfo {author} {\bibfnamefont {Z.}~\bibnamefont {Mao}}, \bibinfo {author}
  {\bibfnamefont {W.}~\bibnamefont {Windl}}, \bibinfo {author} {\bibfnamefont
  {J.}~\bibnamefont {Goldberger}}, \bibinfo {author} {\bibfnamefont
  {M.}~\bibnamefont {Bockrath}}, \ and\ \bibinfo {author} {\bibfnamefont
  {C.~N.}\ \bibnamefont {Lau}},\ }\bibfield  {title} {\enquote {\bibinfo
  {title} {Raman spectroscopy, photocatalytic degradation, and stabilization of
  atomically thin chromium tri-iodide},}\ }\href {\doibase
  10.1021/acs.nanolett.8b01131} {\bibfield  {journal} {\bibinfo  {journal}
  {Nano Letters}\ }\textbf {\bibinfo {volume} {18}},\ \bibinfo {pages}
  {4214--4219} (\bibinfo {year} {2018})}\BibitemShut {NoStop}%
\bibitem [{\citenamefont {Staiger}\ \emph {et~al.}(2015)\citenamefont
  {Staiger}, \citenamefont {Gillen}, \citenamefont {Scheuschner}, \citenamefont
  {Ochedowski}, \citenamefont {Kampmann}, \citenamefont {Schleberger},
  \citenamefont {Thomsen},\ and\ \citenamefont {Maultzsch}}]{RN23}%
  \BibitemOpen
  \bibfield  {author} {\bibinfo {author} {\bibfnamefont {M.}~\bibnamefont
  {Staiger}}, \bibinfo {author} {\bibfnamefont {R.}~\bibnamefont {Gillen}},
  \bibinfo {author} {\bibfnamefont {N.}~\bibnamefont {Scheuschner}}, \bibinfo
  {author} {\bibfnamefont {O.}~\bibnamefont {Ochedowski}}, \bibinfo {author}
  {\bibfnamefont {F.}~\bibnamefont {Kampmann}}, \bibinfo {author}
  {\bibfnamefont {M.}~\bibnamefont {Schleberger}}, \bibinfo {author}
  {\bibfnamefont {C.}~\bibnamefont {Thomsen}}, \ and\ \bibinfo {author}
  {\bibfnamefont {J.}~\bibnamefont {Maultzsch}},\ }\bibfield  {title} {\enquote
  {\bibinfo {title} {Splitting of monolayer out-of-plane
  ${A}_{1}^{\ensuremath{'}}$ {Raman} mode in few-layer \ce{WS2}},}\ }\href
  {\doibase 10.1103/PhysRevB.91.195419} {\bibfield  {journal} {\bibinfo
  {journal} {Phys. Rev. B}\ }\textbf {\bibinfo {volume} {91}},\ \bibinfo
  {pages} {195419} (\bibinfo {year} {2015})}\BibitemShut {NoStop}%
\bibitem [{\citenamefont {Song}\ \emph {et~al.}(2016)\citenamefont {Song},
  \citenamefont {Tan}, \citenamefont {Zhang}, \citenamefont {Wu}, \citenamefont
  {Sheng}, \citenamefont {Wan}, \citenamefont {Wang}, \citenamefont {Dai},\
  and\ \citenamefont {Tan}}]{RN24}%
  \BibitemOpen
  \bibfield  {author} {\bibinfo {author} {\bibfnamefont {Q.~J.}\ \bibnamefont
  {Song}}, \bibinfo {author} {\bibfnamefont {Q.~H.}\ \bibnamefont {Tan}},
  \bibinfo {author} {\bibfnamefont {X.}~\bibnamefont {Zhang}}, \bibinfo
  {author} {\bibfnamefont {J.~B.}\ \bibnamefont {Wu}}, \bibinfo {author}
  {\bibfnamefont {B.~W.}\ \bibnamefont {Sheng}}, \bibinfo {author}
  {\bibfnamefont {Y.}~\bibnamefont {Wan}}, \bibinfo {author} {\bibfnamefont
  {X.~Q.}\ \bibnamefont {Wang}}, \bibinfo {author} {\bibfnamefont
  {L.}~\bibnamefont {Dai}}, \ and\ \bibinfo {author} {\bibfnamefont {P.~H.}\
  \bibnamefont {Tan}},\ }\bibfield  {title} {\enquote {\bibinfo {title}
  {Physical origin of {Davydov} splitting and resonant {Raman} spectroscopy of
  {Davydov} components in multilayer \ce{MoTe2}},}\ }\href {\doibase
  10.1103/PhysRevB.93.115409} {\bibfield  {journal} {\bibinfo  {journal} {Phys.
  Rev. B}\ }\textbf {\bibinfo {volume} {93}},\ \bibinfo {pages} {115409}
  (\bibinfo {year} {2016})}\BibitemShut {NoStop}%
\bibitem [{\citenamefont {Froehlicher}\ \emph {et~al.}(2015)\citenamefont
  {Froehlicher}, \citenamefont {Lorchat}, \citenamefont {Fernique},
  \citenamefont {Joshi}, \citenamefont {Molina-Sanchez}, \citenamefont
  {Wirtz},\ and\ \citenamefont {Berciaud}}]{RN25}%
  \BibitemOpen
  \bibfield  {author} {\bibinfo {author} {\bibfnamefont {G.}~\bibnamefont
  {Froehlicher}}, \bibinfo {author} {\bibfnamefont {E.}~\bibnamefont
  {Lorchat}}, \bibinfo {author} {\bibfnamefont {F.}~\bibnamefont {Fernique}},
  \bibinfo {author} {\bibfnamefont {C.}~\bibnamefont {Joshi}}, \bibinfo
  {author} {\bibfnamefont {A.}~\bibnamefont {Molina-Sanchez}}, \bibinfo
  {author} {\bibfnamefont {L.}~\bibnamefont {Wirtz}}, \ and\ \bibinfo {author}
  {\bibfnamefont {S.}~\bibnamefont {Berciaud}},\ }\bibfield  {title} {\enquote
  {\bibinfo {title} {Unified description of the optical phonon modes in
  {N}-layer \ce{MoTe2}},}\ }\href {\doibase 10.1021/acs.nanolett.5b02683}
  {\bibfield  {journal} {\bibinfo  {journal} {Nano Letters}\ }\textbf {\bibinfo
  {volume} {15}},\ \bibinfo {pages} {6481--6489} (\bibinfo {year}
  {2015})}\BibitemShut {NoStop}%
\bibitem [{\citenamefont {Kim}\ \emph {et~al.}(2016)\citenamefont {Kim},
  \citenamefont {Lee}, \citenamefont {Nam},\ and\ \citenamefont
  {Cheong}}]{RN26}%
  \BibitemOpen
  \bibfield  {author} {\bibinfo {author} {\bibfnamefont {K.}~\bibnamefont
  {Kim}}, \bibinfo {author} {\bibfnamefont {J.-U.}\ \bibnamefont {Lee}},
  \bibinfo {author} {\bibfnamefont {D.}~\bibnamefont {Nam}}, \ and\ \bibinfo
  {author} {\bibfnamefont {H.}~\bibnamefont {Cheong}},\ }\bibfield  {title}
  {\enquote {\bibinfo {title} {Davydov splitting and excitonic resonance
  effects in {Raman} spectra of few-layer \ce{MoSe2}},}\ }\href {\doibase
  10.1021/acsnano.6b04471} {\bibfield  {journal} {\bibinfo  {journal} {ACS
  Nano}\ }\textbf {\bibinfo {volume} {10}},\ \bibinfo {pages} {8113--8120}
  (\bibinfo {year} {2016})}\BibitemShut {NoStop}%
\bibitem [{\citenamefont {Tian}\ \emph {et~al.}(2016)\citenamefont {Tian},
  \citenamefont {Gray}, \citenamefont {Ji}, \citenamefont {Cava},\ and\
  \citenamefont {Burch}}]{Tian2016}%
  \BibitemOpen
  \bibfield  {author} {\bibinfo {author} {\bibfnamefont {Y.}~\bibnamefont
  {Tian}}, \bibinfo {author} {\bibfnamefont {M.~J.}\ \bibnamefont {Gray}},
  \bibinfo {author} {\bibfnamefont {H.}~\bibnamefont {Ji}}, \bibinfo {author}
  {\bibfnamefont {R.~J.}\ \bibnamefont {Cava}}, \ and\ \bibinfo {author}
  {\bibfnamefont {K.~S.}\ \bibnamefont {Burch}},\ }\bibfield  {title} {\enquote
  {\bibinfo {title} {Magneto-elastic coupling in a potential ferromagnetic {2D}
  atomic crystal},}\ }\href {\doibase 10.1088/2053-1583/3/2/025035} {\bibfield
  {journal} {\bibinfo  {journal} {2D Materials}\ }\textbf {\bibinfo {volume}
  {3}},\ \bibinfo {pages} {025035} (\bibinfo {year} {2016})}\BibitemShut
  {NoStop}%
\bibitem [{\citenamefont {Klemens}(1966)}]{RN27}%
  \BibitemOpen
  \bibfield  {author} {\bibinfo {author} {\bibfnamefont {P.~G.}\ \bibnamefont
  {Klemens}},\ }\bibfield  {title} {\enquote {\bibinfo {title} {Anharmonic
  decay of optical phonons},}\ }\href {\doibase 10.1103/PhysRev.148.845}
  {\bibfield  {journal} {\bibinfo  {journal} {Phys. Rev.}\ }\textbf {\bibinfo
  {volume} {148}},\ \bibinfo {pages} {845--848} (\bibinfo {year}
  {1966})}\BibitemShut {NoStop}%
\bibitem [{\citenamefont {Martin}\ and\ \citenamefont {Varma}(1971)}]{RN28}%
  \BibitemOpen
  \bibfield  {author} {\bibinfo {author} {\bibfnamefont {R.~M.}\ \bibnamefont
  {Martin}}\ and\ \bibinfo {author} {\bibfnamefont {C.~M.}\ \bibnamefont
  {Varma}},\ }\bibfield  {title} {\enquote {\bibinfo {title} {Cascade theory of
  inelastic scattering of light},}\ }\href {\doibase
  10.1103/PhysRevLett.26.1241} {\bibfield  {journal} {\bibinfo  {journal}
  {Phys. Rev. Lett.}\ }\textbf {\bibinfo {volume} {26}},\ \bibinfo {pages}
  {1241--1244} (\bibinfo {year} {1971})}\BibitemShut {NoStop}%
\bibitem [{\citenamefont {Merlin}\ \emph {et~al.}(1978)\citenamefont {Merlin},
  \citenamefont {G\"untherodt}, \citenamefont {Humphreys}, \citenamefont
  {Cardona}, \citenamefont {Suryanarayanan},\ and\ \citenamefont
  {Holtzberg}}]{RN29}%
  \BibitemOpen
  \bibfield  {author} {\bibinfo {author} {\bibfnamefont {R.}~\bibnamefont
  {Merlin}}, \bibinfo {author} {\bibfnamefont {G.}~\bibnamefont
  {G\"untherodt}}, \bibinfo {author} {\bibfnamefont {R.}~\bibnamefont
  {Humphreys}}, \bibinfo {author} {\bibfnamefont {M.}~\bibnamefont {Cardona}},
  \bibinfo {author} {\bibfnamefont {R.}~\bibnamefont {Suryanarayanan}}, \ and\
  \bibinfo {author} {\bibfnamefont {F.}~\bibnamefont {Holtzberg}},\ }\bibfield
  {title} {\enquote {\bibinfo {title} {Multiphonon processes in \ce{YbS}},}\
  }\href {\doibase 10.1103/PhysRevB.17.4951} {\bibfield  {journal} {\bibinfo
  {journal} {Phys. Rev. B}\ }\textbf {\bibinfo {volume} {17}},\ \bibinfo
  {pages} {4951--4958} (\bibinfo {year} {1978})}\BibitemShut {NoStop}%
\end{thebibliography}%


%

\clearpage
\newpage
\noindent\textbf{Figure1}

\begin{figure*}[h]
\includegraphics[scale=0.9]{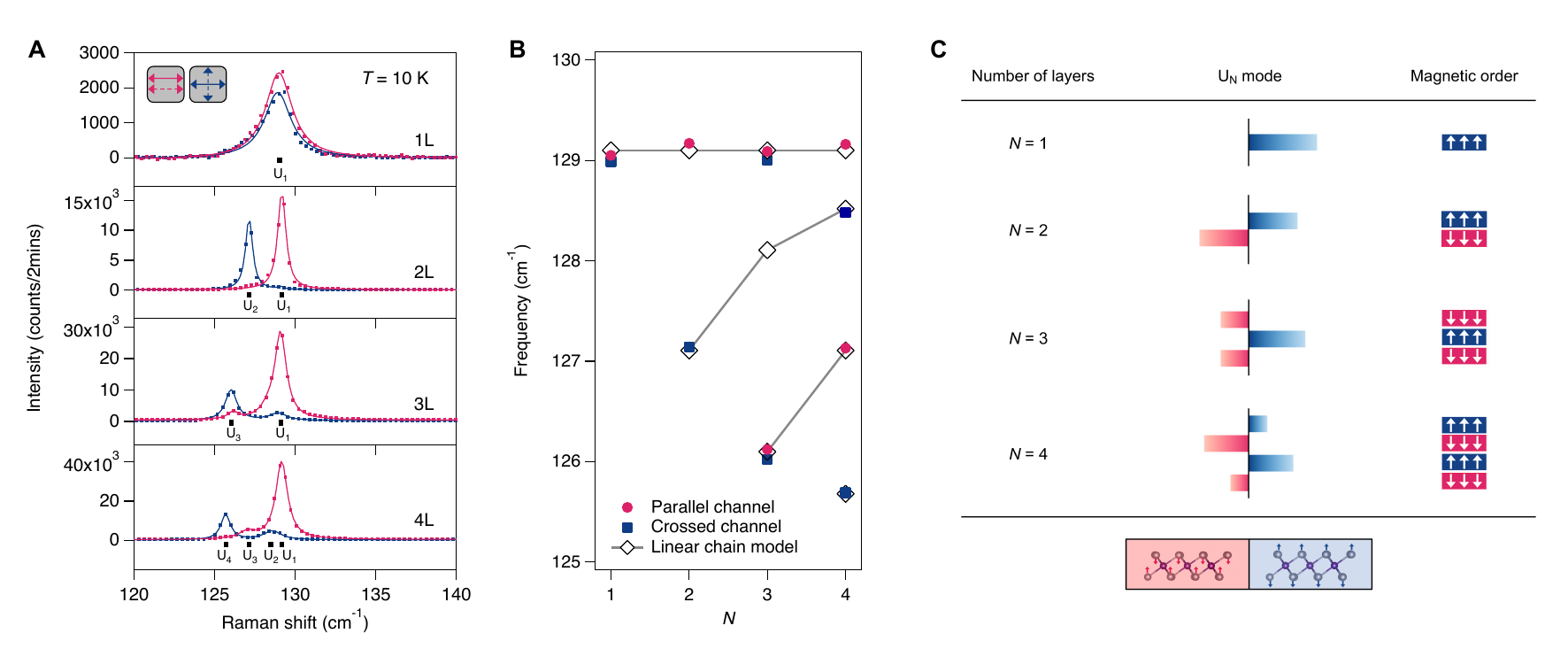}
\caption{\label{fig:fig.1} (A) Raman spectra of 1--4L \ce{CrI3} acquired in linear parallel (ruby) and crossed (royal) channels at 10 K. The solid curves are fits to the raw data (dots). The vertical bars underneath individual spectra denote the fitted frequencies and $U_i (i=1, 2, \cdots,N)$ labels the corresponding modes in $N$-layer \ce{CrI3}. (B) Plot of the fitted frequencies of the modes in (A) as a function of layer number $N$. Ruby filled circles and royal filled squares correspond to modes extracted from the linear parallel and crossed channels, respectively. Solid curves with open diamonds are fits to the Davydov-split frequencies calculated from the linear chain model. (C) Atomic displacement of the lowest frequency mode $U_N$ along with the layered-AFM order to illustrate that $\vec{U}_i\cdot \vec{M}$ maximized at $i = N$ where $\vec{U}_i$ and $\vec{M}$ share the same parity. The rectangular bar represents the atomic displacement amplitude and phase for individual layers, by its length and color (ruby and royal for opposite phase), respectively.}
\end{figure*}

\clearpage
\newpage
\noindent\textbf{Figure2}

\begin{figure*}[h]
\includegraphics[scale=0.9]{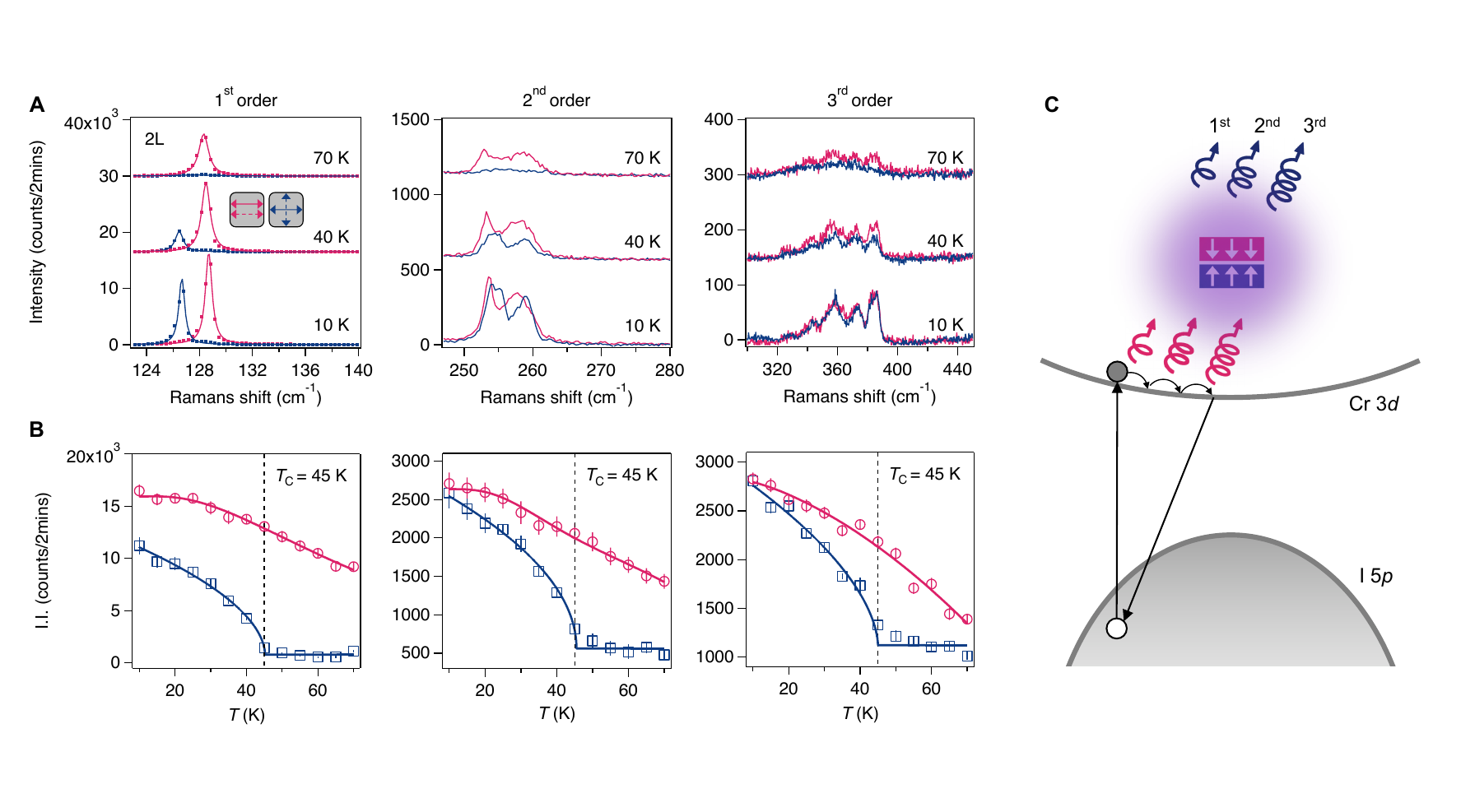}
\caption{\label{fig:fig.2} (A) $1^\mathrm{st}$, $2^\mathrm{nd}$, and $3^\mathrm{rd}$ order Raman spectra of 2L \ce{CrI3} in linear parallel (ruby) and crossed (royal) channels at selected temperatures of 70 K, 40 K, and 10 K. Spectra are vertically offset for clarity. $1^\mathrm{st}$ order spectra show raw data points and fitting curves, and $2^\mathrm{nd}$ and $3^\mathrm{rd}$ spectra are raw spectra. (B) Temperature dependence of integrated intensity (I. I.) of $1^\mathrm{st}$, $2^\mathrm{nd}$, and $3^\mathrm{rd}$ order Raman modes in the parallel (ruby circles) and crossed (royal squares) channels. Solid curves are fits to the anharmonic decay model in the parallel channel (ruby curves) and the order parameter-like function $I_0+I\sqrt{T_\mathrm{C}-T}$ in the crossed channel (royal curves). Critical temperature $T_\mathrm{C}$ = 45 K is marked by a dashed vertical line in each panel. (C) Schematic illustration of the cascading process of phonon scattering (ruby) and its layered-AFM assisted counterpart (royal). The springs with one to three windings represent $1^\mathrm{st}$ to $3^\mathrm{rd}$ order process.}
\end{figure*}

\clearpage
\newpage
\noindent\textbf{Figure3}
\begin{figure*}[h]
\includegraphics[scale=0.9]{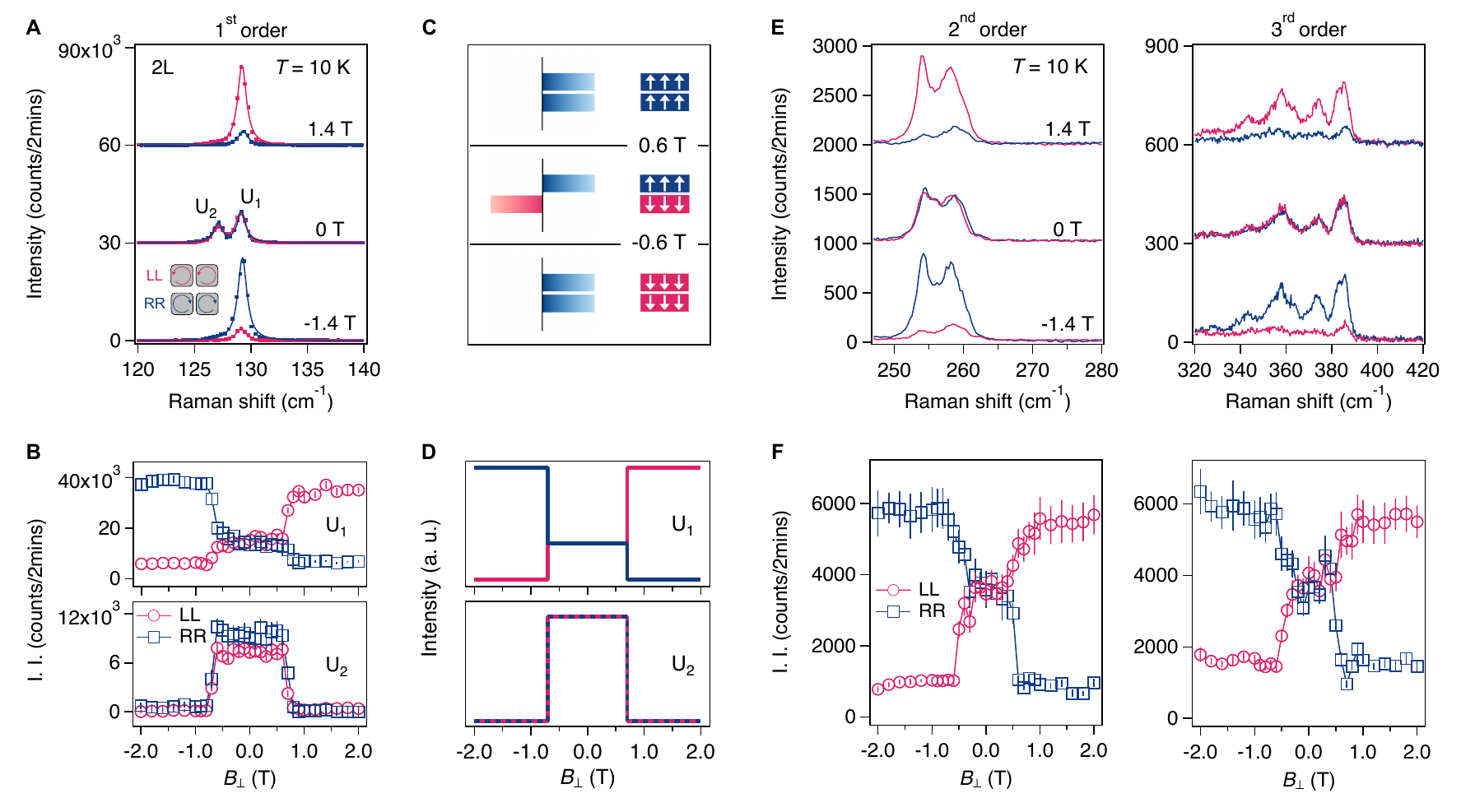}
\caption{\label{fig:fig.3} (A) Raman spectra of 2L \ce{CrI3} in co-circularly polarized channels, LL (ruby) and RR (royal), in selected out-of-plane magnetic field ($B_\bot$) of 1.4 T, 0 T, and -1.4 T. Dots are raw data points and solid curves are fitting curves. (B) Magnetic field dependence of integrated intensity (I. I.) of the two modes of 2L \ce{CrI3}, $U_1$ and $U_2$, in both LL (ruby circles) and RR (royal squares) channels. (C) List of phonon modes in 2L \ce{CrI3} that have finite coupling strength $\vec{U}_i\cdot \vec{M}$ for individual magnetic orders that appear at $B_\bot$ above and between $B_\mathrm{c} = \pm0.6$ T, the critical magnetic field for the layered-AFM to FM transition. (D) Calculated $B_\bot$ dependence of $U_1$ and $U_2$ of 2L \ce{CrI3}. (E) $2^\mathrm{nd}$ and $3^\mathrm{rd}$ order Raman modes (raw spectra) acquired in the same condition as (A). (F) $B_\bot$ dependence of I. I. of $2^\mathrm{nd}$ and $3^\mathrm{rd}$ order modes of 2L \ce{CrI3}.}
\end{figure*}

\clearpage
\newpage
\noindent\textbf{Figure4}
\begin{figure*}[h]
\includegraphics[scale=0.9]{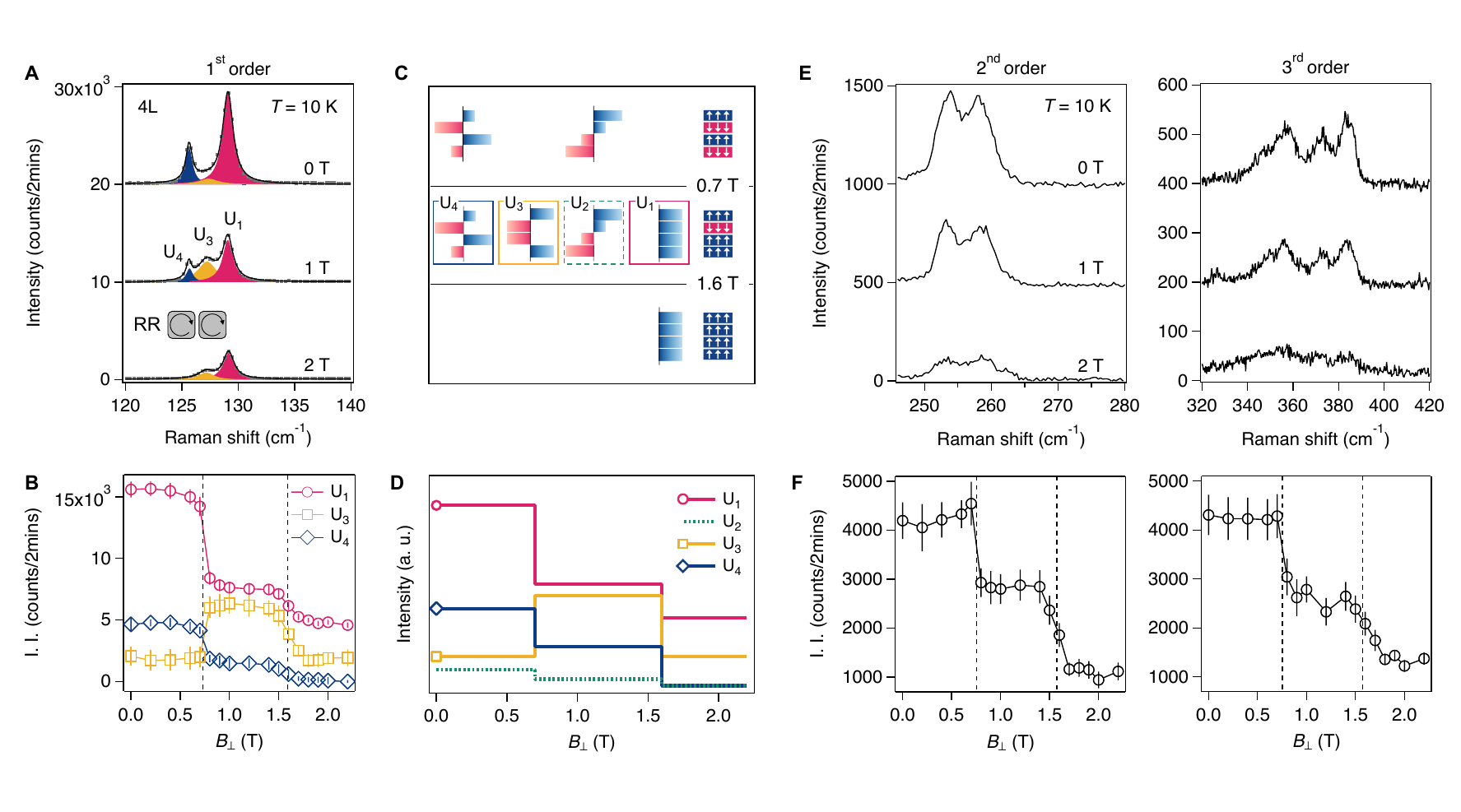}
\caption{\label{fig:fig.4} (A) Raman spectra of 4L \ce{CrI3} in the RR channel in selected $B_\bot$ of 0 T, 1 T and 2 T. Grey dots are raw data points and black curves are fits to multiple Lorentzian function. Lorentzian profiles in ruby, honey, and royal correspond to modes $U_1$, $U_3$, and $U_4$ of 4L \ce{CrI3}. (B) $B_\bot$ dependence of I. I. of $U_1$ (ruby circles), $U_3$ (honey squares), and $U_4$ (royal diamonds). Vertical dashed lines mark the critical magnetic fields for spin flip transitions, $B_\mathrm{c1}$ = 0.7 T and $B_\mathrm{c2}$ = 1.6 T. (C) List of phonon modes in 4L \ce{CrI3} that have finite coupling strength $\vec{U}_i\cdot \vec{M}$ for magnetic orders below $B_\mathrm{c1}$, between $B_\mathrm{c1}$ and $B_\mathrm{c2}$, and above $B_\mathrm{c2}$. Dash emerald box marks $U_2$ that is not observed in (A) due to its weak intensity and closeness to $U_1$. Solid boxes of ruby, honey, and royal color highlight $U_1$ (ruby), $U_3$ (honey), and $U_4$ (royal), respectively. (E) and (F) $B_\bot$ dependence of I. I. of $2^\mathrm{nd}$, and $3^\mathrm{rd}$ order modes of 4L \ce{CrI3}.}
\end{figure*}

\clearpage
\newpage
\noindent \textit{Supplementary Material}\\
\begin{center}
\vspace{12pt}
\textbf{\large Tunable layered-magnetism-assisted rich magneto-Raman effect in a two-dimensional magnet  \ce{CrI3}}\\

Wencan Jin$,^{1, \ast}$ Zhipeng Ye$,^3$ Xiangpeng Luo$,^1$ Bowen Yang$,^2$ Gaihua Ye$,^3$ Hyun Ho Kim$,^{2, \dagger}$ Fangzhou Yin$,^2$ Laura Rojas$,^3$ Shangjie Tian$,^4$ Hechang Lei$,^4$ Adam W. Tsen$,^2$ Kai Sun$,^1$ Rui He$^{3, \ddagger}$ and Liuyan Zhao$,^{1, \S}$\\

\textit{$^1$Department of Physics, University of Michigan, 450 Church Street,\\ Ann Arbor, Michigan 48109, USA}\\

\textit{$^2$Institute for Quantum Computing, Department of Chemistry, \\and Department of Physics and Astronomy, University of Waterloo,\\ Waterloo, 200 University Ave W, Ontario N2L 3G1, Canada}

\textit{$^3$Department of Electrical and Computer Engineering, 910 Boston Avenue, \\Texas Tech University, Lubbock, Texas 79409, USA}

\textit{$^4$Department of Physics and Beijing Key Laboratory of \\Opto-electronic Functional Materials \& Micro-nano Devices, \\Renmin University of China, Beijing 100872 China}
\end{center}

\vspace{20pt}

\noindent \textbf{Table of Contents}\\
\noindent{I. Full-range Raman spectra for 1--4L \ce{CrI3}}\\
\noindent{II. Davydov splitting of $A_\mathrm{g}$ mode in $N$-layer \ce{CrI3} described by a linear chain model}\\
\noindent{III. Magnetic field dependence of $U_2$ and $U_1$ modes in 2L \ce{CrI3}}\\
\noindent{IV. Magnetic field dependence of $U_{1-4}$ modes in 4L \ce{CrI3}}\\

\vspace{0.75 in}

\begin{footnotesize}
\noindent{$\ast$ current affiliation: Department of Physics, Auburn University, 380 Duncan Drive, Auburn, AL 36849, USA}\\
\noindent{$\dagger$ current affiliation: School of Materials Science and Engineering, Kumoh National Institute of Technology, Gumi, Gyeongbuk 39177, Korea}\\
\noindent{$\ddagger$ rui.he@ttu.edu}\\
\noindent{$\S$ lyzhao@umich.edu}
\end{footnotesize}


\newpage
\noindent\textbf{I. Full-range Raman spectra for 1--4L \ce{CrI3}}

\begin{figure*}[h]
\includegraphics[scale=0.75]{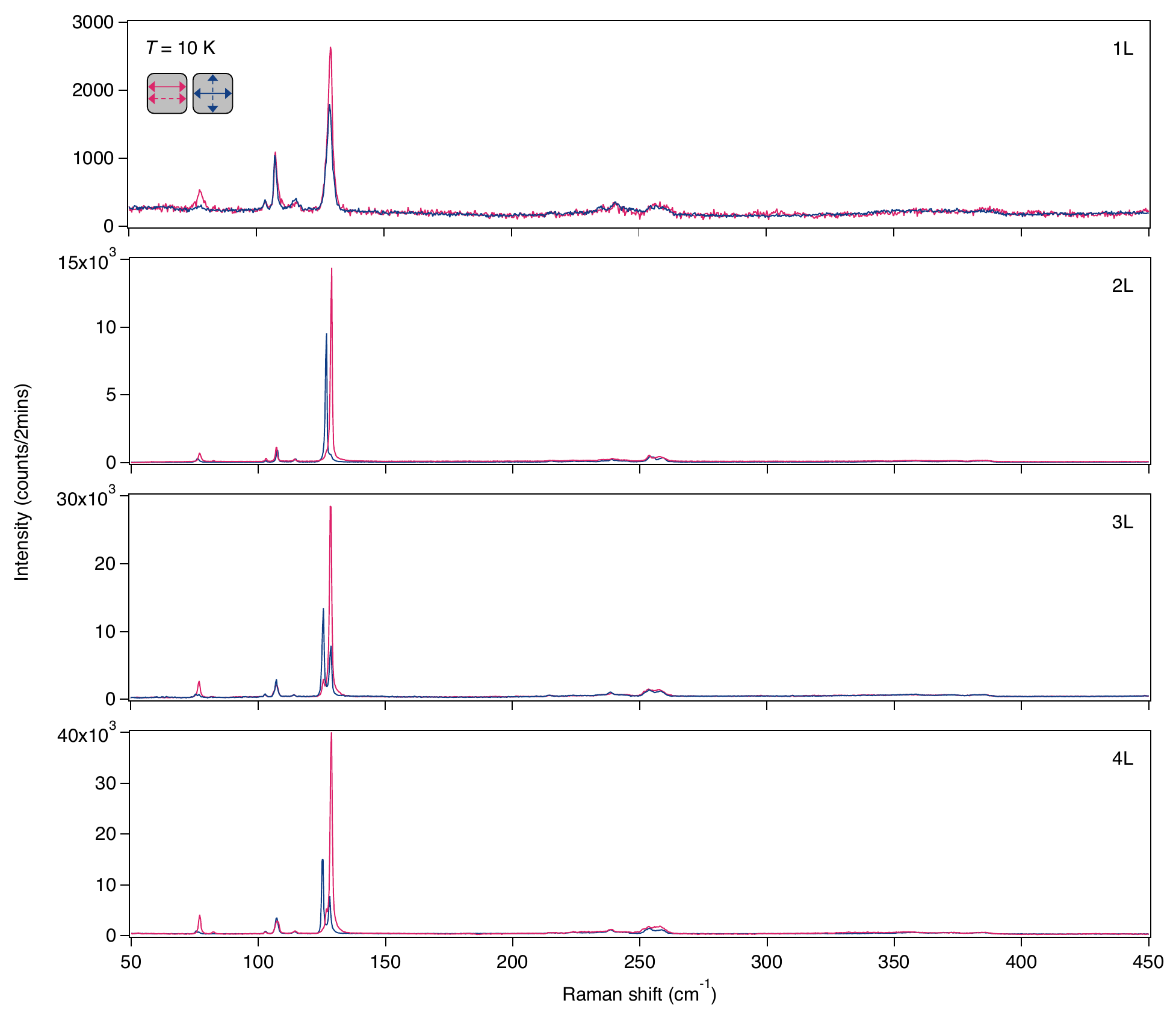}
\end{figure*}
\vspace{-20pt}
\begin{footnotesize}
\noindent \textbf{Fig. S1.} Full range Raman spectra of 1-4L \ce{CrI3} in the linear parallel and crossed channels acquired at 10 K using 633 nm lasers.\\
\end{footnotesize}

\noindent We notice that 1L \ce{CrI3} has a linewidth of $\sim2.2 \mathrm{cm}^{-1}$, clearly broader than those of 2L, 3L, and 4L \ce{CrI3} ($\sim$0.6-1.0 $\mathrm{cm}^{-1}$). Note that in reference Nature Nanotech. 15, 212 (2020) and Nano Letters 20, 729 (2020), a broad linewidth ($\sim$4.0 $\mathrm{cm}^{-1}$ and $\sim$3.5 $\mathrm{cm}^{-1}$, respectively) for 1L \ce{CrI3} was observed and no explanation has been provided.\\

\noindent We think it may result from a few factors: (a) the ultimate 2D limit of 1L \ce{CrI3} may have stronger fluctuations than its thicker counterparts. (b) the interfacing with hBN flakes on both sides of a \ce{CrI3} flake makes 1L \ce{CrI3} fully in contact with hBN that could cause strong damping effect to phonons. Such an effect is weaker in thicker layers because of their lower surface-to-volume ratio. Or (c) there are simply more defects introduced to 1L \ce{CrI3} as it is much harder to achieve 1L \ce{CrI3} than its thicker counterparts.\\

\noindent The main purpose for using the 633 nm resonant excitation is to enhance the signal level and achieve a good signal to noise ratio. Meanwhile, this resonant excitation does not alter the Raman selection rule. As shown in Fig. S2, Raman spectra taken in both parallel and crossed channels on a same 2L \ce{CrI3} flake with 633 nm and 532 nm excitations exhibit no observable selection rule difference, but a significant enhancement of signal level (by a factor of $\sim$50) with the 633 nm laser.

\begin{figure*}[h]
\includegraphics[scale=0.6]{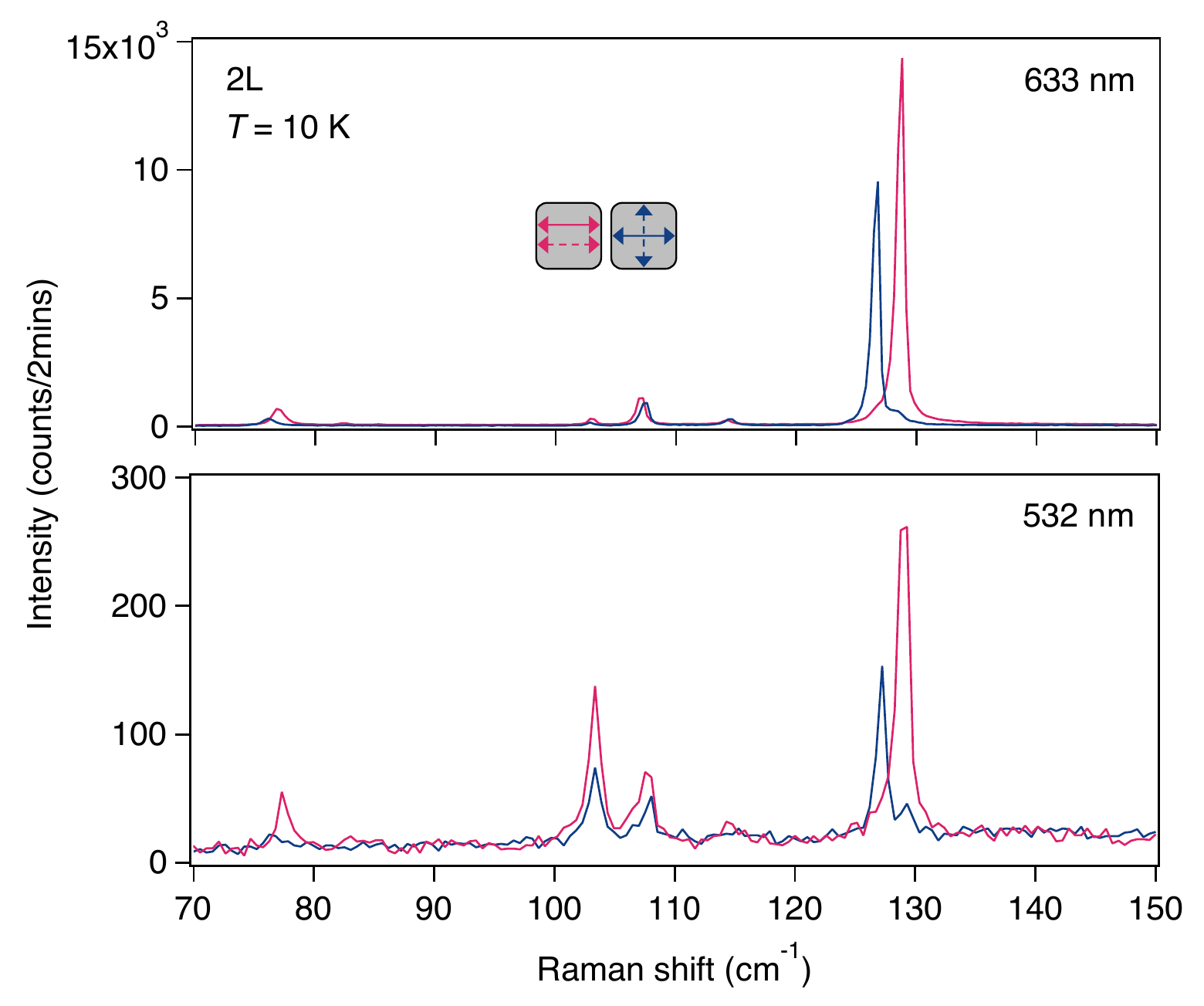}
\end{figure*}
\vspace{-20pt}
\begin{footnotesize}
\noindent \textbf{Fig. S2.} Raman spectra of 2L \ce{CrI3} in the linear parallel and crossed channels using 633 nm and 532 nm lasers.\\
\end{footnotesize}


\newpage
\noindent\textbf{II. Davydov splitting of $A_\mathrm{g}$ mode in $N$-layer \ce{CrI3} described by a linear chain model}\\

\noindent The Davydov splitting of $A_\mathrm{g}$ mode in $N$-layer \ce{CrI3} can be accounted for by a linear chain model written as 
\begin{center}
$H=\sum_{i=1}^{N}(\frac{1}{2}m\dot{u}_i^2+\frac{1}{2}k_0u_i^2)+\frac{1}{2}\sum_{i=2}^{N}k(u_{i-1}-u_i)^2=\frac{1}{2}U^T KU$
\end{center}
\vspace{10pt}

\noindent with eigenmode $U=\begin{pmatrix} u_1 \\ u_2 \\ \vdots \\ u_N \end{pmatrix}$, in which $u_i$ contains both amplitude and phase of the atomic displacement within the $i^\mathrm{th}$ layer; $k_0$ and $k$ stand for the force constant of the original $A_\mathrm{g}$ mode of monolayer \ce{CrI3} and the coupling constant between adjacent layers, respectively. 

\noindent We thus obtain a $N\times N$ matrix $K={\begin{pmatrix} k_0+k & -k & 0 & \cdots & \cdots & 0 \\ -k & k_0+2k & -k &  &  & 0 \\ 0 & -k & k_0+2k &  &  & 0 \\ \vdots &  &  & \ddots &  & \vdots \\ \vdots &  &  &  & k_0+2k & -k \\ 0 & 0 & 0 & \cdots & -k & k_0+k\end{pmatrix}}_{N\times N} = k_0 I + kG$, where $I$ is an identify matrix, and $G$ matrix describes interlayer coupling. Diagonalizing the matrix $G$ gives rise to the $N$ eigenvalues ($e_i$ with $i = 1, 2, \cdots, N$) and eigenmodes for $N$-layer \ce{CrI3} and their parity (see Table S1). We can also obtain the corresponding eigenfrequencies $\Omega(\omega_0, \omega)=\sqrt{\omega_0^2+e_i\cdot \omega^2}$, with $\omega_0 = \sqrt{\frac{k_0}{m}}$, and $\omega=\sqrt{\frac{k}{m}}$ (see Table S2).\\

\noindent\textbf{Absence of $U_2$ mode in the spectrum of 3L \ce{CrI3}}

\noindent As shown in Table S1, $U_2$ mode of 3L \ce{CrI3} $U_2=\frac{1}{\sqrt{2}} \begin{pmatrix} 1\\ 0  \\ -1\end{pmatrix}$ is parity odd, which is Raman silent in the parallel channel. Meanwhile, the coupling between $U_2$ with layered AFM order $M=\begin{pmatrix} 1\\-1\\1\end{pmatrix}$ also vanishes. As a result, it cannot be observed in the crossed channel either. Therefore, $U_2$ mode is absent in our spectrum of 3L \ce{CrI3} shown in Fig. 1A. 

\newpage
\noindent \textbf{Table. S1.} $N$ eigenmodes for $N$-layer \ce{CrI3} and their parities. 
\begin{figure*}[h]
\includegraphics[scale=0.62]{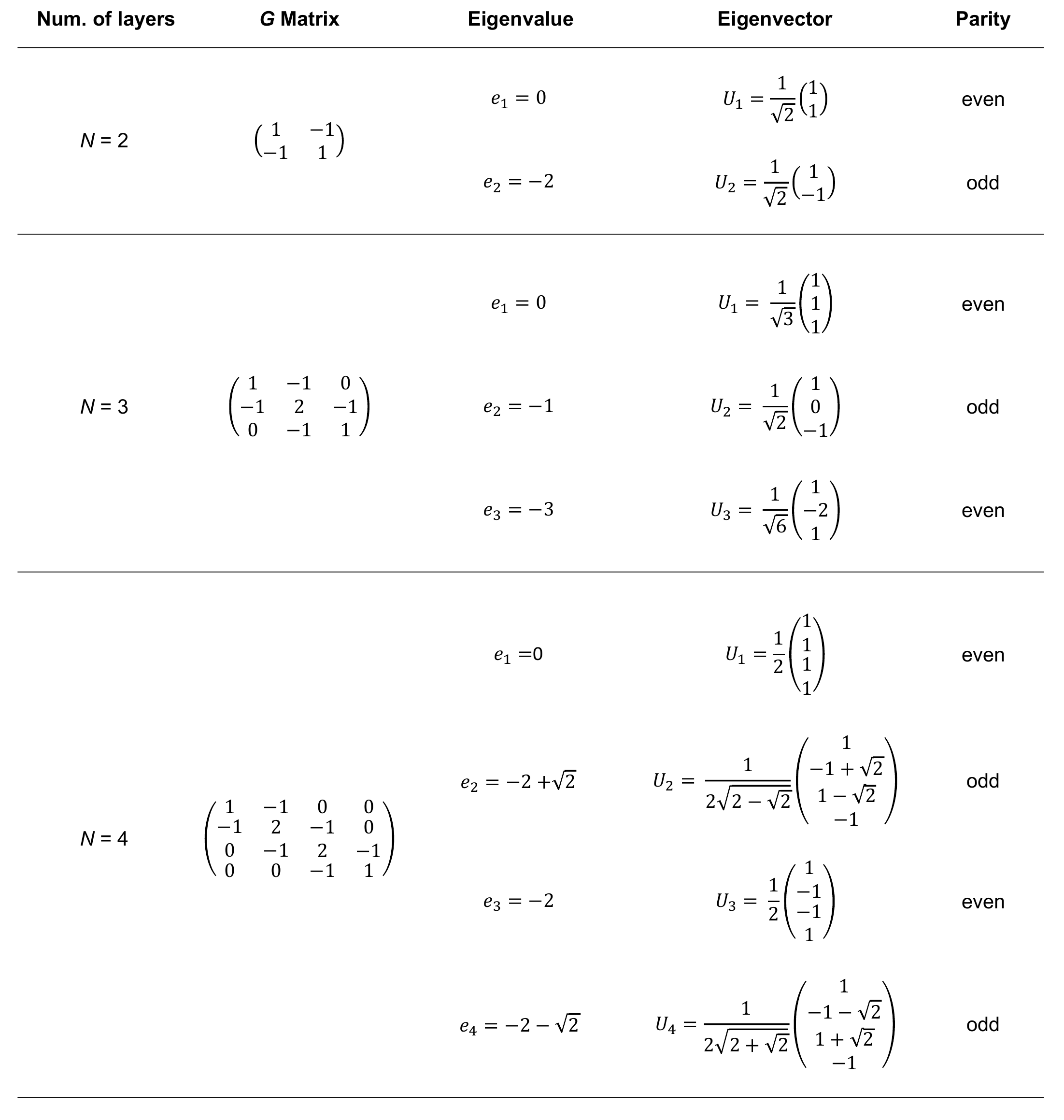}
\end{figure*}

\noindent \textbf{Table. S2.} Functional forms of $N$ eigenfrequencies for $N$-layer \ce{CrI3}. 
\begin{figure*}[h]
\includegraphics[scale=0.62]{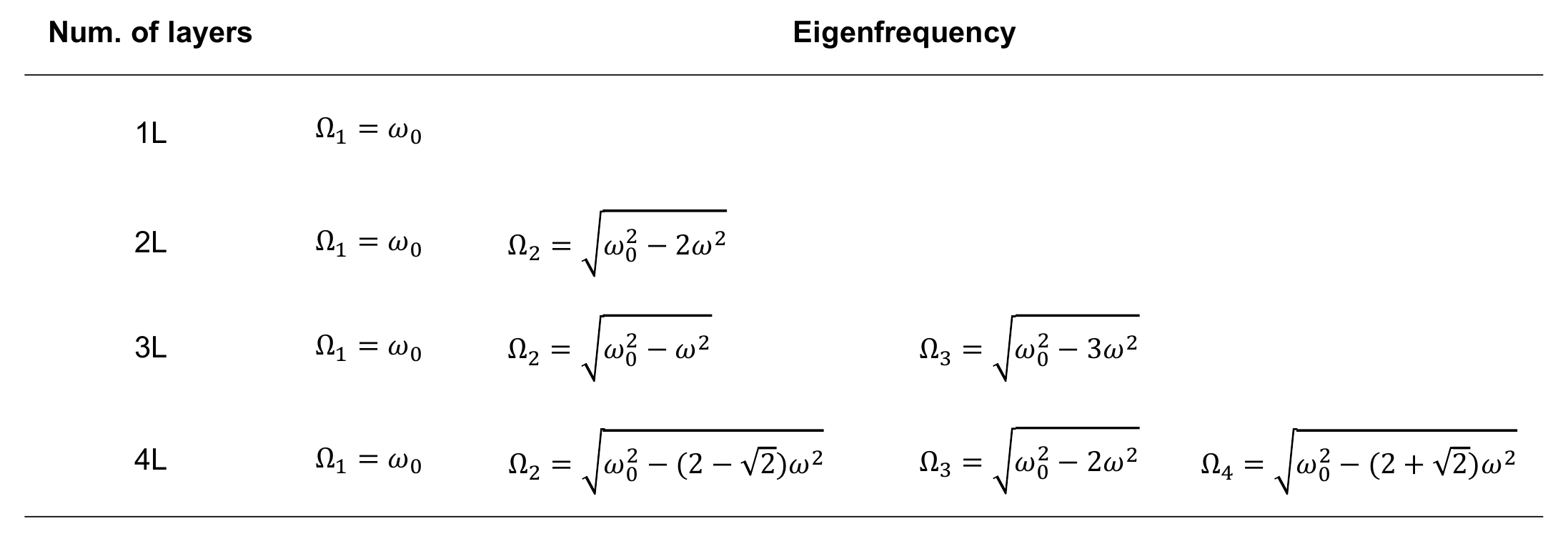}
\end{figure*}

\newpage
\noindent \textbf{Table. S3.} Computed Raman selection rules and intensities for the layered-AFM state in $N$-layer \ce{CrI3}.
\begin{figure*}[h]
\includegraphics[scale=0.54]{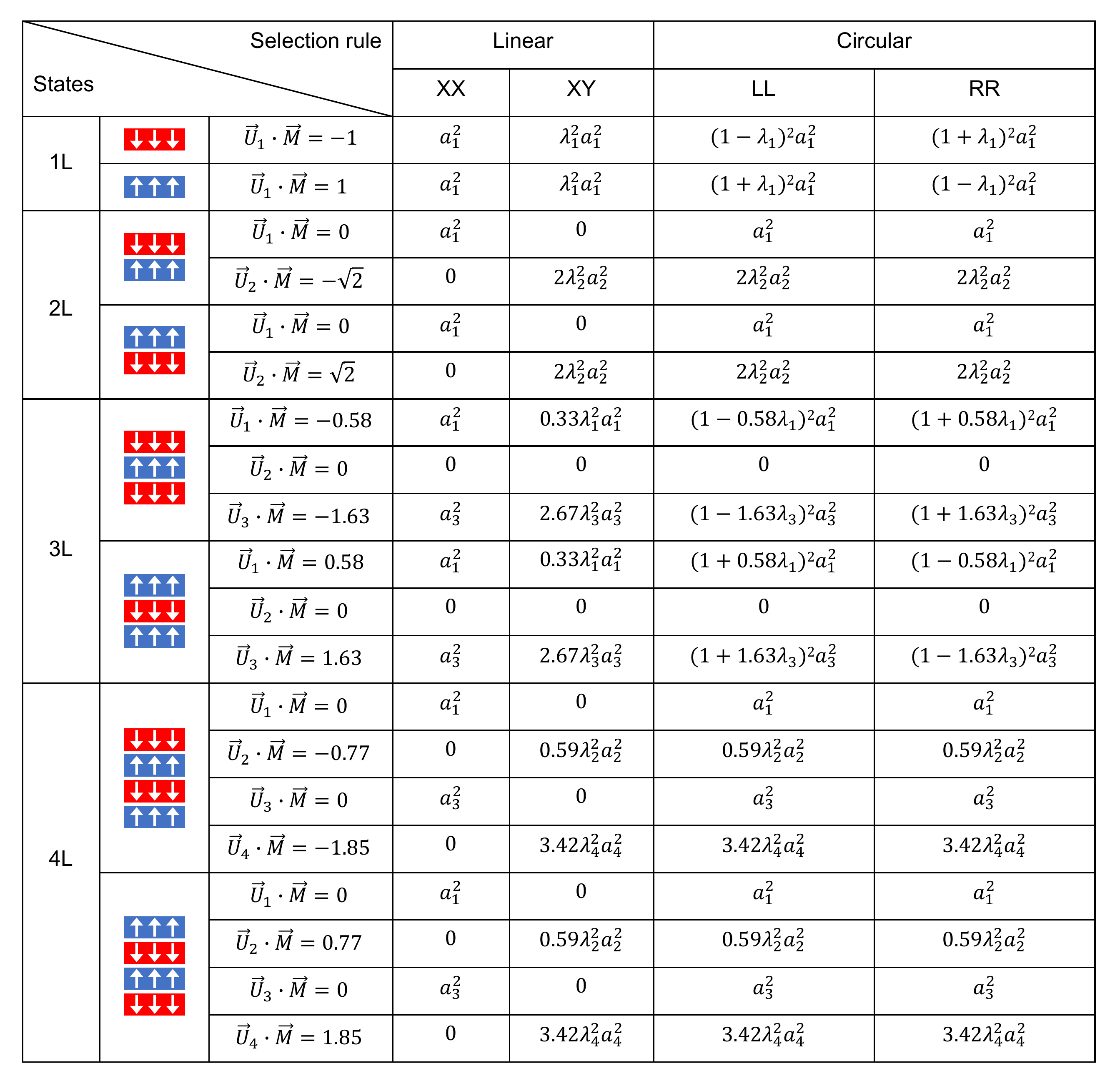}
\end{figure*}

\begin{figure*}[h]
\includegraphics[scale=0.6]{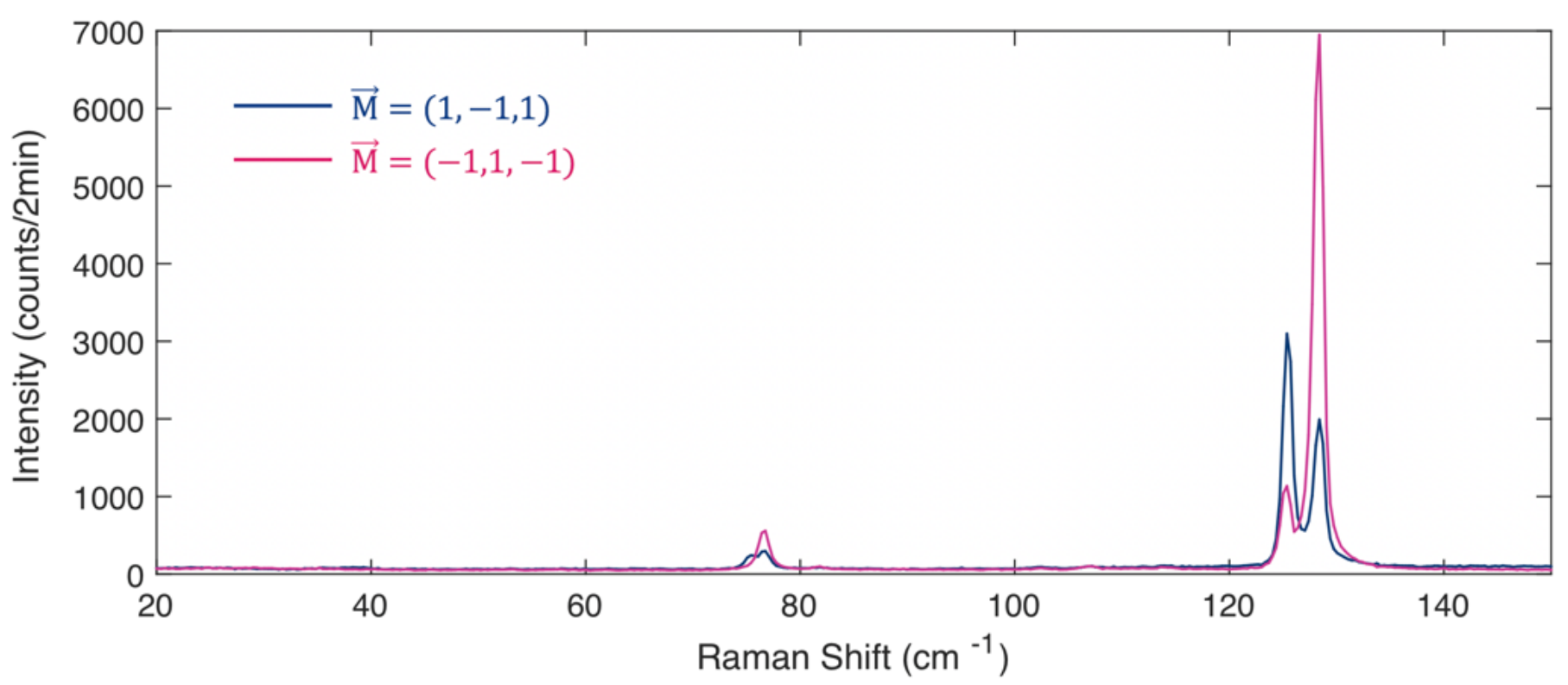}
\end{figure*}
\vspace{-20pt}
\begin{footnotesize}
\noindent \textbf{Fig. S3.} Raman spectra of 3L \ce{CrI3} in the RR channel for the two time-reversal symmetry related magnetic ground states $\vec{M}$ = (1,-1,1) and (-1,1,-1).\\
\end{footnotesize}


\newpage
\noindent\textbf{III. Magnetic field dependence of $U_2$ and $U_1$ modes in 2L \ce{CrI3}}\\

\noindent\textbf{Table. S4.} Coupling strength ($\vec{U}_i\cdot \vec{M}$) between the phonon mode ($\vec{U}_i$) with the magnetic order ($\vec{M}$).

\begin{figure*}[h]
\includegraphics[scale=0.62]{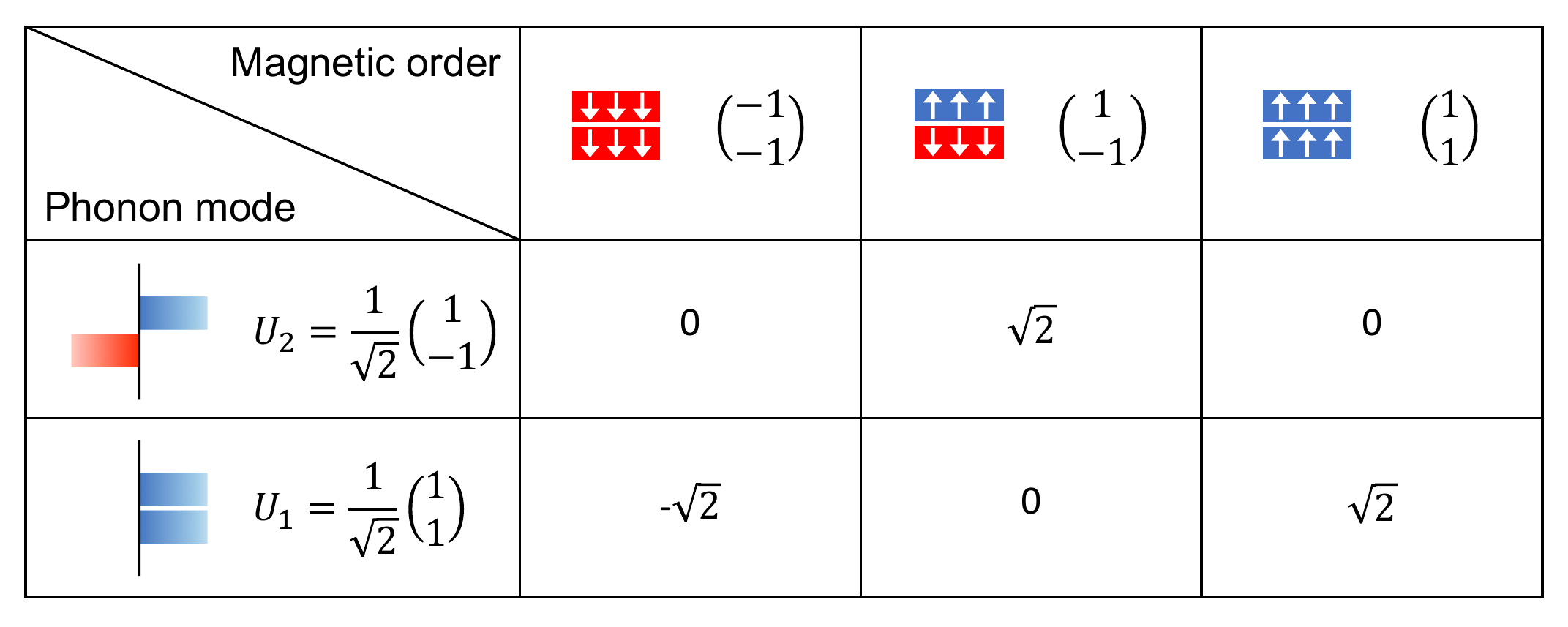}
\end{figure*}

\noindent\textbf{Table. S5.} Calculated Raman tensor forms and selection rules for the two modes in 2L \ce{CrI3}.
\begin{figure*}[h]
\includegraphics[scale=0.75]{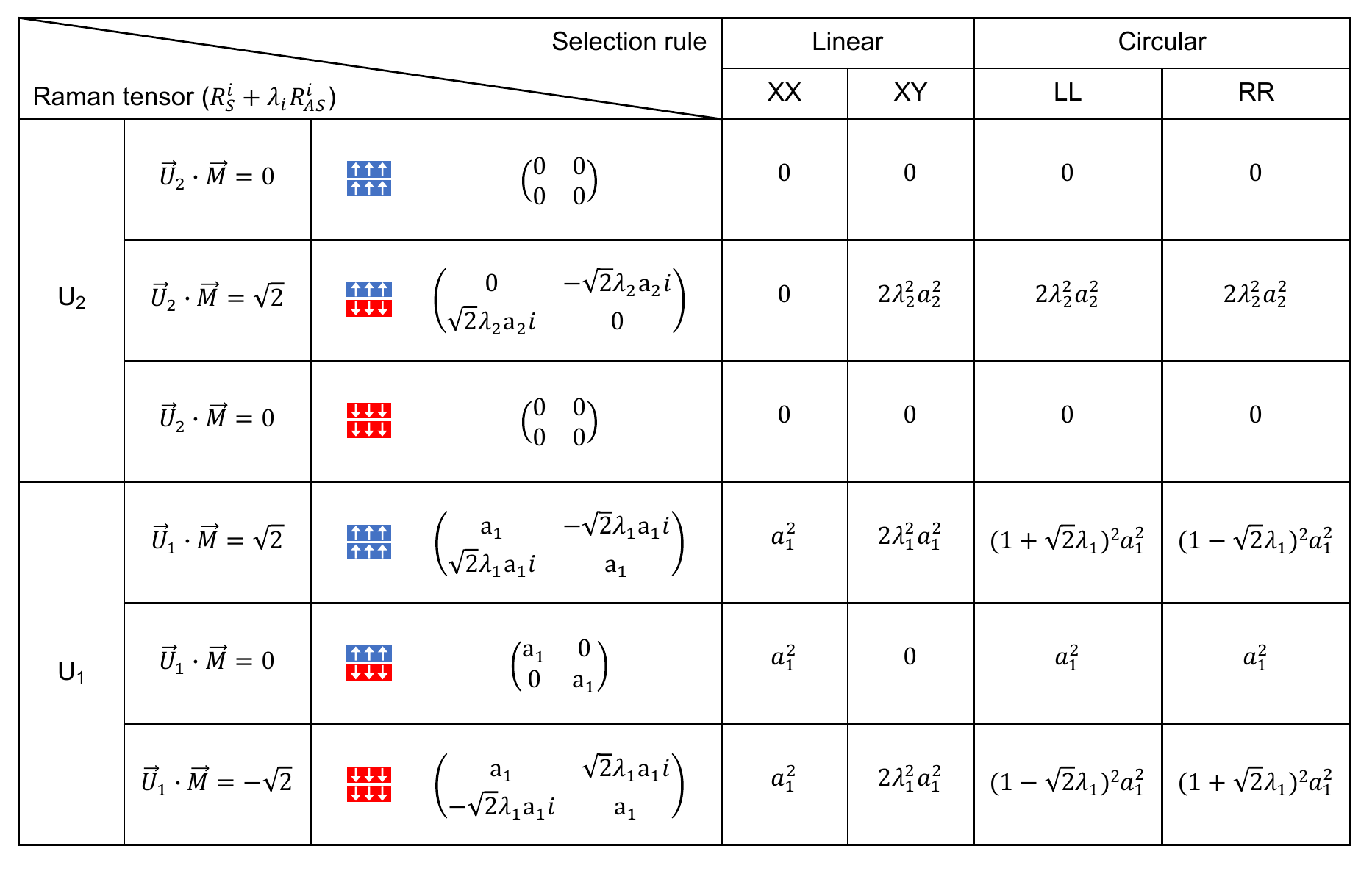}
\end{figure*}


\newpage
\noindent\textbf{IV. Magnetic field dependence of $U_{1-4}$ modes in 4L \ce{CrI3}}\\

\noindent\textbf{Table. S6.} Coupling strength ($\vec{U}_i\cdot \vec{M}$) between the phonon mode ($\vec{U}_i$) with the magnetic order ($\vec{M}$).

\begin{figure*}[h]
\includegraphics[scale=0.75]{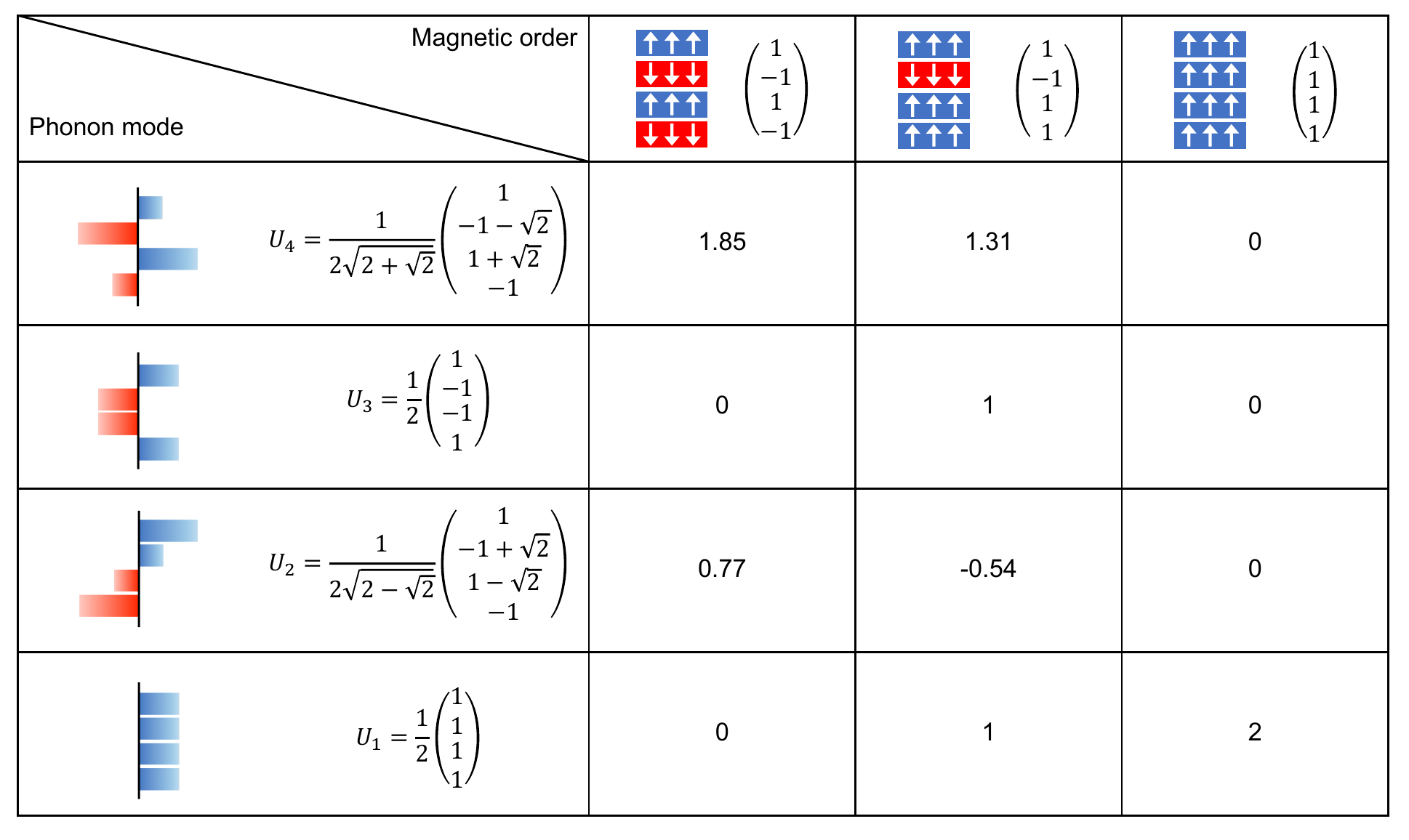}
\end{figure*}

\newpage
\noindent\textbf{Table. S7.} Calculated Raman tensor forms and selection rules for the two modes in 4L \ce{CrI3}.
\begin{figure*}[h]
\includegraphics[scale=0.55]{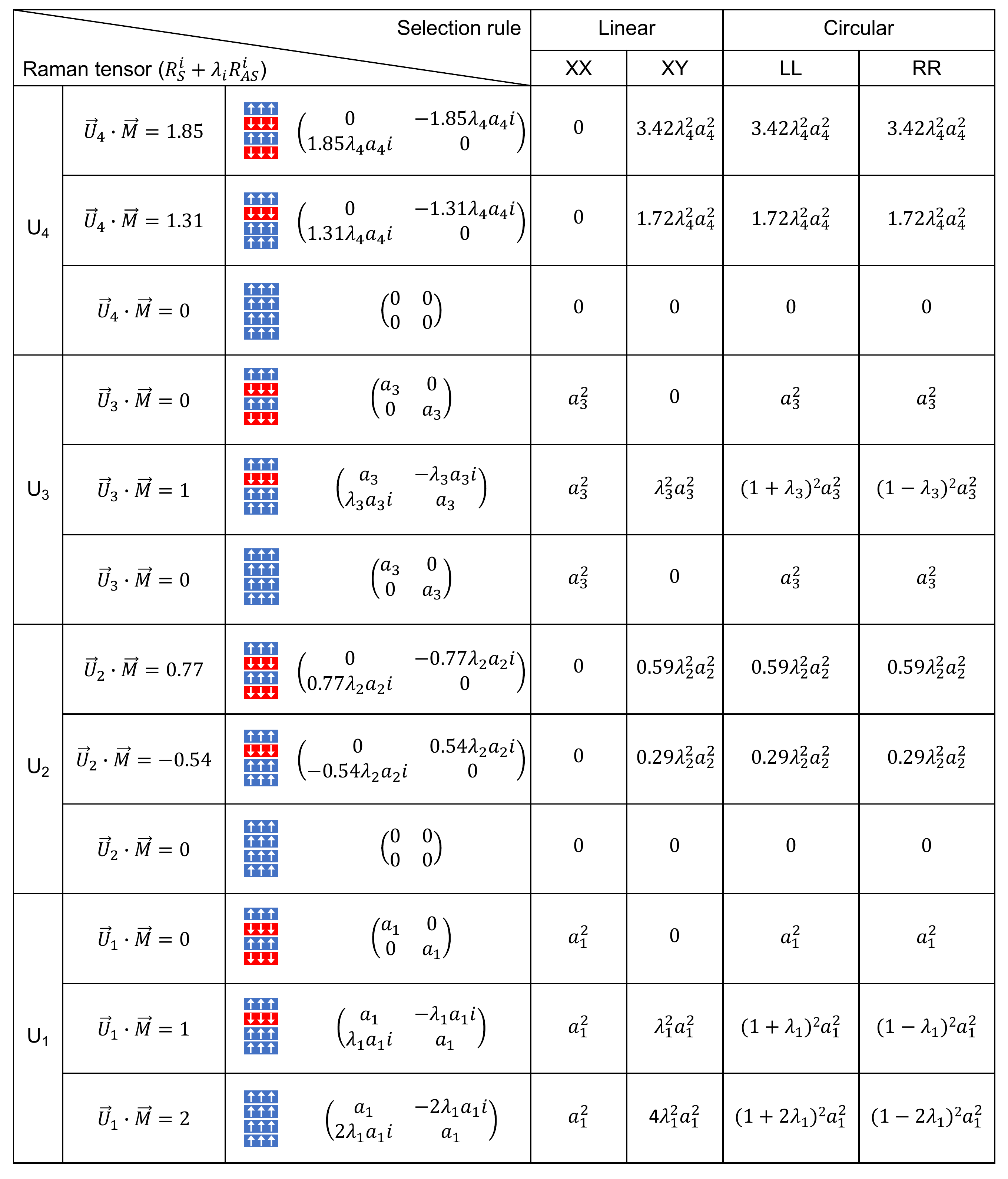}
\end{figure*}

\end{document}